\documentclass[letterpaper,english,aps, prb, superscriptaddress, eprint, longbibliography, floatfix, twocolumn, balancelastpage, flushbottom]{revtex4-2}
\usepackage{xcolor}
\usepackage{babel}
\usepackage{verbatim}
\usepackage{amsmath}
\usepackage{amssymb}
\usepackage{graphicx}
\PassOptionsToPackage{normalem}{ulem}
\usepackage{ulem}
\usepackage[pdfusetitle,
 bookmarks=true,bookmarksnumbered=false,bookmarksopen=false,
 breaklinks=true,pdfborder={0 0 0},pdfborderstyle={},backref=false,colorlinks=true,
 citecolor=blue]
 {hyperref}
\usepackage[all]{hypcap}
\usepackage{orcidlink} 

\makeatletter

\pdfpageheight\paperheight
\pdfpagewidth\paperwidth

\providecommand{\tabularnewline}{\\}

\makeatother

\begin{document}

\title{Transparent Graphene-Superconductor Interfaces:\\Quantum Hall and Zero Field Regimes}

\author{Alexey Bondarev}
\affiliation{Department of Physics, Duke University, P.O. Box 90305, Durham, North
Carolina 27708-0305, USA}
\author{Gu Zhang\,\orcidlink{0000-0003-1073-1029}}
\altaffiliation[current address: ]{School of Physics, Nanjing University, China} 
\affiliation{Beijing Academy of Quantum Information Sciences, Beijing 100193, China}
\author{Harold U. Baranger\,\orcidlink{0000-0002-1458-2756}}
\email{harold.baranger@duke.edu}
\affiliation{Department of Physics, Duke University, P.O. Box 90305, Durham, North
Carolina 27708-0305, USA}

\date{9 June 2025; \href{https://doi.org/10.1103/gjcm-qfgz}{Phys. Rev. B 111, 235444 (2025) Editors' Suggestion} (accepted version)}

\begin{abstract}
We study clean, edge-contacted graphene/superconductor interfaces in both the quantum Hall (QH) and zero field regimes.  We find that Andreev reflection is substantially stronger than at an interface with a semiconductor two-dimensional electron gas: the large velocity at graphene's conical Dirac points makes the requirement of current continuity to a metal much less restrictive.  
In both our tight-binding and continuum models, we find a wide range of parameters for which Andreev reflection is strong.  For a \emph{transparent} interface, we demonstrate the following for graphene in the lowest Landau level QH state: (i)~Excellent electron-hole hybridization occurs: the electron and hole components in graphene are simply related by an exchange of sublattice.  
The spatial profile for the electron component is predominantly gaussian on one sublattice and peaked at the interface, and so very different from the QH edge state of a terminated lattice.  (ii)~The degree of hybridization is independent of the graphene filling: no fine-tuning is needed.  
(iii)~The spectrum is valley degenerate: the dispersion of each chiral Andreev edge mode (CAEM) self-aligns to be antisymmetric about the center of each valley, independent of filling.  Achieving a transparent interface requires the absence of any barrier as well as a superconductor that is suitably matched to graphene; we argue that the latter condition is not very stringent.  
We further consider the effect of reduced transparency and Zeeman splitting on both the wavefunctions of the CAEM and their dispersion. 
\end{abstract}


\maketitle

\section{Introduction \label{sec:Intro}}

\subsection{Motivation and Context}

The interface between graphene and a superconductor has attracted great interest for nearly two decades from several points of view (for reviews see Refs.\,\citep{BeenakkerRMP08,LeeLeeReviewRPP18}). First, there has been a long-standing desire for semiconductor-superconductor structures that are gateable and so controllable \citep{vanWees_Sem-SupReview_chapter97,DohDeFranceschi_TunableIS_Sci05}, leading potentially to 
interesting quantum devices  
\citep{AkazakiJJ_APL96, WenShabani_LogicIEEE19, PhanShabani_PampPRAppl23, Huang-GraphDetect-X24}. 
By encapsulating graphene between two cladding layers (typically BN), such a high mobility and 
gateable structure can be made from graphene \citep{DeanHone_BN_NatNano10,MayorovGeim_Encapsulate_NL11,WangDean_1DcontactSci13, Young_QHferromag-Gr_NatPhy12, CaladoDelftNatNano15}. 

Graphene has a Dirac spectrum at low energy, and so a second motivation has been to investigate how the superconducting proximity effect is manifest in such a material. The Dirac spectrum results from an additional degree of freedom, namely the sublattice A or B in the unit cell of the honeycomb lattice, which can be treated as a Dirac spinor. There are two inequivalent Dirac cones in graphene's bandstructure, referred to as valleys and labeled $K$ and $K'$. 
At low energy, this valley degree of freedom can also be treated as a spinor. Thus, the quantum state of graphene may exhibit SU(4) symmetry, which on the other hand may be broken 
\citep{Young_QHferromag-Gr_NatPhy12}.
It has been of interest to study how these degrees of freedom interact with a superconductor in proximity \citep{CaladoDelftNatNano15,BenShalomManchesterNatPhy16, FrancoisGlebSci16}. 

The quantum Hall (QH) effect can be seen in graphene at low magnetic field, of order $1\,$T. Since the upper critical field of several thin-film superconductors is of order $10\,$T, interfaces between graphene in the QH state and several s-wave superconductors have been made \citep{CaladoDelftNatNano15,BenShalomManchesterNatPhy16, FrancoisGlebSci16,ParkLee_S-QHedge_SciRep17, LeeHarvardNatPhy17, SahuBangalorePRL18, VignaudSacepe_ChiralSuper_Nat23}. This, then, is an interface between very different quantum states: the BCS state of the superconductor with a gap $\Delta$ and the topological QH state of graphene with a gap of order the inter-Landau-level energy $E_{L}$. 
Since both systems are gapped, there are no bulk propagating states at the chemical potential. 
However, there is no obvious gap along the interface, and in fact there are propagating modes confined to the graphene-superconductor interface. We focus on such interface modes, which we call chiral Andreev edge modes (CAEM). 

Finally, there has been great interest in interfaces between topological and superconducting states because they can host exotic interface modes \citep{AliceaRPP12,Beenakker_Search_Rev13,QiHuguesZhangPRB10}. For instance, Majorana modes produced in this way \citep{QiHuguesZhangPRB10} could provide a basis for topological quantum computation \citep{MongUniversalPRX14,LianZhangChiralMajPNAS18}. 
The QH/superconductor interface is the simplest interface combining topological and superconducting states, and so provides a first step in this direction. 

For all of these reasons there has been substantial work on graphene-superconductor interfaces, starting very soon after the discovery of graphene \citep{BeenakkerRMP08,LeeLeeReviewRPP18}. 
A key step was the development of contacts to encapsulated graphene on the edge of the graphene sheet \citep{WangDean_1DcontactSci13}. This allowed for high-quality samples with one-dimensional graphene/superconductor contact  \citep{CaladoDelftNatNano15,BenShalomManchesterNatPhy16,FrancoisGlebSci16, IvanGlebBallisticPRL16,ParkLee_S-QHedge_SciRep17,LeeHarvardNatPhy17, SahuBangalorePRL18, VignaudSacepe_ChiralSuper_Nat23,ParkLeePRL18}, which is the geometry that we consider here. 
QH edge and interface modes were observed: a variety of quantum transport properties were studied including the conductance and gating of the CAEM \citep{ParkLee_S-QHedge_SciRep17,LeeHarvardNatPhy17, SahuBangalorePRL18, SeredinskiGlebSideGateSciAdv19,LingfeiGlebLandButPRB24}, 
interference among them \citep{LingfeiGlebCAESNatPhys20,LingfeiGlebLossPRL23}, 
and electrical noise associated with them \citep{SahuDas_Noise_PRB21,LingfeiGlebThermalPRL25}. 
In addition the supercurrent in Josephson junctions in which QH graphene is placed between two superconductors has been studied  
\citep{FrancoisGlebSci16,ZhuBenShalom_EdgeShunt_NatCom17,TingGlebJJsPRR19,VignaudSacepe_ChiralSuper_Nat23}. 
Edge-contacted graphene/superconductor has clearly proved to be an excellent, controlled system; its main competitor in QH/super\-conductor interface physics is the InAs/super\-conductor system with extended planar contact between the two materials, with which a number of similar quantum transport experiments have been done \citep{EromsAndreevPRL05,Kozuka_SQH-Oxide_JPSJ18, MatsuoSciRep18, HatefipourShabani_QH-S_NL22, HatefipourShabani_Andreev-QH-QPC_PRB24}. 

Theoretical interest in the QH/superconductor system predates the recent experiments \citep{MaZyuzin_S-QH_EPL93, Zyuzin_S-QH-S_PRB94,TakagakiQHEPRB98,HoppeZulickePRL00,Zulicke_SupSemHighB_Physica01, GiazottoPRB05, AkhmerovValleyPolarPRL07, KhaymovichEPL10, StoneLinPRB11, VanOstaayTripletPRB11, LianEdgestatePRB16} but has certainly increased because  of them. In particular, recent papers \citep{GamayunCheianovPRB17,AlaviradSauPRB18,BeconciniNonlocalCorr_PRB18, Manesco-QHgrS_SP22, Peralta-Bariloche_EdgeTran_PRB21,TangAlicea_VortexEnabled_PRB22, Galambos-CAR-PRB22, KurilovichDisorderAES_NCom23,MichelsenSchmidt_SupercurEnable_PRR23, DavidGrenoble_Geometrical_PRB23,SchillerOreg_FermDissip_PRB23,Khrapai_BogoliubovInterferom_PRB23, CuozzoRossiPRB24,HuLian-Decohere-X24} 
addressed, for instance, interference among  interface modes, effects of sample geometry and disorder, and the roles of dissipation and supercurrent in QH/superconductor hybridization, though not necessarily in the context of graphene. 

Our goal in this work is to understand the properties of infinite, uniform graphene/superconductor interfaces in the QH regime.
Because the experimental interfaces are thought to be highly transparent \citep{LingfeiGlebCAESNatPhys20,ParkLeePRL18}, we consider interfaces in which the electrons in graphene are coupled strongly to those in the superconductor: there should be no interfacial barrier and the wavefunctions should match well (explained further below).  
We refer to this situation loosely by saying the interface is ``transparent''.  
For such interfaces, we find that the special properties of graphene lead to new behavior. 


\subsection{Summary of Main Results}

We present results for a clean interface between a zigzag edge of graphene in the QH regime and a superconductor, using both continuum and tight-binding models. In both cases we solve the single-particle Bogoliubov-de Gennes equation (BdG) that describes inhomogeneous, mean-field superconductors \citep{deGennesBook,BeenakkerRMP08}. 
In the continuum model, we match wavefunctions between a Dirac spectrum material and one with parabolic dispersion with a large Fermi energy. In the tight-binding approach, we stitch together a honeycomb lattice for the graphene with a square lattice near half filling for the superconductor. 
The two main assumptions are that (i) there is no disorder in either material or at the interface and (ii) the fields in the BdG equation (the pair potential, one-body potential, and magnetic field) all change abruptly at the graphene-superconductor interface.  

Andreev reflection is  
a convenient way to view the proximity effect in many situations \citep{deGennesBook,vanWees_Sem-SupReview_chapter97,Klapwijk_ProximityAndreev_2004,Asano_AndreevBook21}. 
In this approach, one focuses on the excitations on the normal side of the interface. These can be either an electron above the Fermi sea or the absence of an electron below the Fermi sea, namely a hole. Individual excitations cannot propagate into the superconductor; however, two electrons together can connect to the (nearly) zero momentum condensate in the superconductor. 
In Andreev reflection, an incoming electron with wavevector $k$ at the interface 
pairs with a second electron at $-k$ and goes into the superconducting condensate. The system therefore has a deficit of $-k$, which is described by an outgoing hole with wavevector $k$.
Thus, the superconductor couples electrons and holes with the same momentum (at the chemical potential). 
In Andreev reflection in graphene, then, an electron in the $K$ valley reflects as a hole also in the $K$ valley.  The sublattice structure of this hole is the same as that of the missing $-k$ electron. 
Note that this is one of the conventional terminologies 
\citep{deGennesBook,ColemanBook,Asano_AndreevBook21}
but not the only one in general use 
\citep{BTK-PRB82,Klapwijk_ProximityAndreev_2004,BeenakkerRMP08,DavidGrenoble_Geometrical_PRB23}.  

The degree of electron-hole hybridization in the mode along the interface is a natural way to characterize the strength of Andreev scattering at high field. Equal weight of electrons and holes in each CAEM means that the proximity effect is maximal (strong Andreev reflection), 
while weak Andreev reflection leads to modes that are predominately either electron or hole.  
Quantitatively, we define the ``fractional electron content in graphene,'' $f_e$, by integrating over the electron and hole density in graphene:
\begin{align}
f_e \equiv \frac{\sum_\textrm{sublatt.}\int_{-\infty}^0 dy |\psi^{e}(y)|^2}
{\sum_\textrm{sublatt.}\int_{-\infty}^0 dy \left[ |\psi^{e}(y)|^2 + |\psi^{h}(y)|^2 \right]}, 
\label{eq:fe-define} 
\end{align}
where $\psi^{e}$  ($\psi^{h}$)  is the electron (hole) component of the wavefunction. The setup here is that graphene occupies the lower half-plane ($y<0$) while the superconductor is in the upper.  
The density on both sublattices (A and B) is included; for lattice models, the integral is replaced by a sum on sites. 

\begin{figure*}
\includegraphics[width=\linewidth]{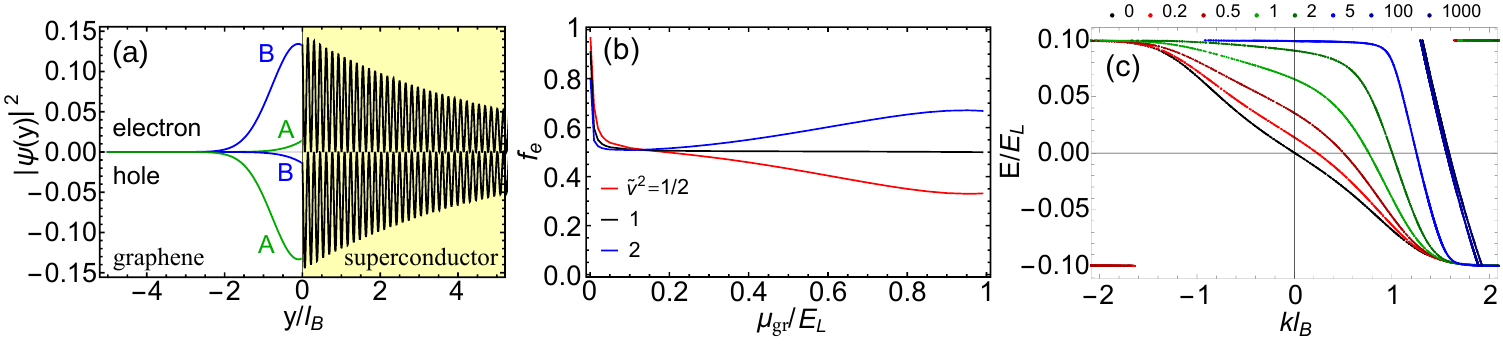}
\caption{Transparent interface between a superconductor ($y>0$) and quantum Hall graphene in the LLL ($y<0$) in the continuum model. 
\textbf{(a) Wavefunction} of the interface mode, $|\psi(y)|^{2}$, in the $K$ valley: the electron and hole components are plotted as \emph{positive} and \emph{negative} numbers, respectively, and in graphene the sublattice A (B) component is green (blue). In graphene the electron and hole components have the same magnitude, indicating perfect Andreev reflection, but the sublattices are exchanged. 
\textbf{(b) Electron-hole hybridization} expressed as the fractional electron content in graphene [Eq.\,(\ref{eq:fe-define})] as a function of the LLL edge state filling ($E_{L}\equiv\sqrt{2}\hbar v/l_{B}$). Shown for three values of the superconductor Fermi energy [cf.\ Eq.~\eqref{eq:def-vtilde} for the definition of $\tilde{v}^2$]. Excellent hybridization is obtained throughout the LLL (except very near the Dirac point). 
\textbf{(c)~Dispersion} of the chiral edge mode ($E<\Delta$) for different strengths of the interface barrier (varying $Z=V_{0}/\hbar v$ to color-coded values).  The spectrum evolves from the terminated-lattice edge mode on the right to the fully-hybridized interface mode centered on the $K$ point ($kl_{B}=0$). 
(The discontinuities in the spectrum are unrelated to the interface state \cite{Discont-Fig1c}.) 
For a transparent interface, then, the spectrum is valley degenerate. 
(Zigzag edge, $\Delta=E_L/10$, for other parameters see Sec.\,\ref{subsec:Parameters}.)
\label{fig:MainResults}}
\end{figure*} 

Three main conclusions emerge from our study, supported by both continuum and tight-binding results. In this introduction, we use results from the continuum model to briefly illustrate each conclusion in 
Fig.\,\ref{fig:MainResults}. 

First, for a transparent interface, we find strong Andreev reflection---in 
the quantum Hall regime as well as at zero magnetic field. 
Current continuity at the interface implies that the ratio of velocities, $v/v_{FS}$, is an important parameter (where $v$ is the velocity at the Dirac point in graphene and $v_{FS}$ is the Fermi velocity in the superconductor) and that one should have $v/v_{FS}\!\sim\!1$ for good matching. Since $v\!\approx\!10^{6}$\,m/sec independent of the graphene chemical potential, achieving this condition is straightforward---unlike for a semiconductor with parabolic dispersion in which the velocity is typically small \cite{v-parabolic}. We find a wide range of parameters that yield excellent Andreev reflection.  

In the QH regime, strong Andreev reflection is illustrated in Fig.\,\ref{fig:MainResults}(a) which shows the transverse wavefunction of the quasi-one-dimensional state at the graphene-superconductor interface in the lowest Landau level (LLL). On the graphene side ($y<0$), apart from a swap of sublattices, the electron and hole densities are nearly identical, showing that Andreev reflection is strong. 
Further discussion of the transverse wavefunction appears in Sec.\,\ref{subsec:Good-eh-Hybrid}, including for non-ideal cases in \ref{sec:Vary-E_Fsup}. 

Second, we show in Fig.\,\ref{fig:MainResults}(b) that this strong electron-hole hybridization is essentially independent of the chemical potential in graphene. In sharp contrast to the case of a semiconductor with parabolic dispersion 
\citep{BTK-PRB82,BlonderTinkhamPRB83,NakanoBTKinterferomPRB93,MortensenAngleAndreevPRB99}, 
for a strongly coupled graphene/superconductor interface, strong Andreev reflection is ubiquitous---no fine-tuning is needed. The upper (lower) curve in Fig.\,\ref{fig:MainResults}(b) corresponds to increasing (decreasing) the Fermi energy of the \emph{superconductor} by a factor of two: such a large change in the superconductor causes only a modest change in electron-hole hybridization in graphene. 
Electron-hole hybridization is discussed further in Secs.\,\ref{subsec:Good-eh-Hybrid} and \ref{sec:Vary-E_Fsup}. 

Third, for a well-matched and transparent interface, the dispersion of the CAEM within the gap of the superconductor is centered on the $K$ point ($k\!=\!0$).
The curves in Fig.\,\ref{fig:MainResults}(c) correspond to changing the magnitude of a barrier between the graphene and superconductor: for $Z\!=0$ there is no interfacial barrier, while $Z\gtrsim5$ is the tunneling regime. The dispersion evolves from the well-studied QH edge mode for a large barrier (curves on the right) to a mode centered on $k\!=0$  for a completely transparent interface ($Z\!=0$, black curve) \cite{Discont-Fig1c}. 
Indeed, for \emph{all} transparent, well-matched interfaces that we studied (see Sec.\,\ref{sec:Dispersion} for further discussion), we find that (i)~the dispersion is at the center of the valley at the chemical potential, $E(k\!=\!0)\!=\!0$, and (ii)~the dispersion is the same in the $K$ and $K'$ valleys---it is ``valley degenerate''. 

The paper is organized as follows. In Sec.\,\ref{sec:Models}, we introduce the specific continuum and tight-binding models, discuss the approximations, and give our typical values for the model parameters. Sec.\,\ref{sec:StrongAndreev} focuses on strong Andreev reflection for transparent interfaces, first at $B=0$ and then in the QH regime. 
In Sec.\,\ref{sec:Dispersion}, we discuss the dispersion of the interface modes and the question of valley degeneracy for both perfect and imperfect interfaces. 
We argue that valley degeneracy, strong Andreev reflection, and strong coupling are naturally connected (Sec.\,\ref{subsec:ValleyDegen-Transp-AR}). The dependence on properties of the superconductor, in particular its chemical potential, are discussed in Sec.\,\ref{sec:Vary-E_Fsup}. 
Though the spin degree of freedom is un\-important in most of our work, in Sec.\,\ref{sec:Spin-Zeeman} we discuss the Zeeman effect, showing how it breaks perfect hybridization but does not break valley degeneracy.  
Implications for Majorana modes are discussed. Finally, our conclusions are in Sec.\,\ref{sec:Conclusions}:  
they include a discussion connecting these infinite-interface results (in particular, $f_e$) to experiments, 
which are typically performed in nanostructures.

\section{Continuum and Tight-Binding Models of the Interface \label{sec:Models}}

We approach the graphene-superconductor interface using the standard inhomogeneous mean field theory of superconductivity. The single-particle excitations of the system are conveniently obtained from the Bogoliubov--de Gennes (BdG) equation \citep{deGennesBook,BeenakkerRMP08,Asano_AndreevBook21}. One doubles the degrees of freedom in the system by introducing holes and writes a single-particle real-space hamiltonian, $\mathcal{H}$, in this doubled (Nambu) space. Concretely, let $H(\mathbf{r})$ be the normal state hamiltonian with one-body potential $U(\mathbf{r})$ and magnetic field described by the vector potential $A(\mathbf{r})$, and let $\Delta(\mathbf{r})$ be the superconducting pairing potential (the gap). Then the BdG hamiltonian is 
\begin{align}
\mathcal{H} & =\left(\begin{array}{cc}
H(\mathbf{r}) & \Delta(\mathbf{r})\\
\Delta^{*}(\mathbf{r}) & -H^{*}(\mathbf{r})
\end{array}\right),
\label{eq:BdG}
\end{align}
where energy is measured from the chemical potential.
The doubling builds in (artificial) particle-hole symmetry due to redundancy: if the wavefunction $(\psi^{e}(\mathbf{r}),\psi^{h}(\mathbf{r}))$ is a solution with energy $E$, then $(-\psi^{h*}(\mathbf{r}),\psi^{e*}(\mathbf{r}))$ is a solution with energy $-E$. In the mean-field BdG equation there are three fields that should be found self-consistently \citep{deGennesBook}: the superconducting gap $\Delta(\mathbf{r})$, the one-body potential $U(\mathbf{r})$, and the vector potential $A(\mathbf{r})$.

\subsection{Abrupt Interface: Magnetic, Pairing, and Electrostatic Fields \label{subsec:AbruptInterface}}

We make a number of important simplifications with respect to the experimental conditions. First, we take the superconductor to be two dimensional (2D) so that the entire system is 2D. In addition, there is no disorder.  

Furthermore, we assume that the self-consistent fields in the BdG approach 
change abruptly at the graphene-superconductor interface, as commonly done in the literature 
\citep{BTK-PRB82, TakagakiQHEPRB98,HoppeZulickePRL00, Zulicke_SupSemHighB_Physica01, KhaymovichEPL10, AlaviradSauPRB18, BeconciniNonlocalCorr_PRB18, Manesco-QHgrS_SP22,DavidGrenoble_Geometrical_PRB23,CuozzoRossiPRB24}. 
The one-body potential $U(\mathbf{r})$ certainly changes abruptly when the material changes. We neglect a possible long-range  electrostatic contribution to $U(\mathbf{r})$ caused by charge transfer from the superconductor to graphene.
However, the sudden change in the periodic potential is included through a discontinuity in the Fermi energy.  Here, $\mu_{\textrm{gr}}$ is the chemical potential in graphene measured from the Dirac point, while $\mu_S$ for the superconductor is measured from the bottom of the band. $\mu_{\textrm{gr}}$ and $\mu_S$ thus denote the Fermi energies as well, and we have   
 $U(\mathbf{r}) \!=\! -\mu_{\textrm{gr}}$ within graphene and $=\! -\mu_S$ within the superconductor [recall that energy is measured from the chemical potential in Eq.\,(\ref{eq:BdG})].

The pairing interaction vanishes in graphene, implying that $\Delta(\mathbf{r})\!=\!0$.  Because the density of states in graphene is much smaller than in the superconductor, the absence of pairing interaction in graphene has little effect on the superconductor. Therefore, in the superconductor we take the magnitude of the gap, $|\Delta|$, to be constant. Though $\Delta(\mathbf{r})$ switches off abruptly at the interface, pairing correlations extend into the graphene, leading to electron-hole hybridization.  

The magnetic field, $B$, must be large enough to create the quantum Hall state but smaller than the critical field of the superconductor.  The superconductor naturally acts to screen the penetration of the magnetic field.  With this in mind, we assume that $B$ drops abruptly from a constant value in graphene to zero in the superconductor, a simplifying approximation that has been used extensively in the literature \citep{TakagakiQHEPRB98, HoppeZulickePRL00, Zulicke_SupSemHighB_Physica01, KhaymovichEPL10, BeconciniNonlocalCorr_PRB18, Manesco-QHgrS_SP22, DavidGrenoble_Geometrical_PRB23, CuozzoRossiPRB24}. 
The abrupt drop of $B$ implies that the superconductor 
strongly screens the field, as in bulk type I superconductors. 
Though the recent experiments involve type II superconductors \citep{LingfeiGlebCAESNatPhys20, SahuDas_Noise_PRB21, HatefipourShabani_QH-S_NL22}, 
results from the simple abrupt drop approximation yield general insight into QH-superconductor interfaces.  
Other work has made a very different simplifying approximation, namely to take $B$ constant across the interface with no decay \citep{GiazottoPRB05, KurilovichDisorderAES_NCom23, MichelsenSchmidt_SupercurEnable_PRR23}. 

These approximations taken together lead to an enormous simplification of the problem: rather than solving the mean field theory self-consistently, one simply solves the BdG equation once for the fixed, constant values of the potentials: $\Delta$, $B$, and the Fermi energies $\mu_{\textrm{gr}}$ and $\mu_S$.
The result is perhaps an overly optimistic picture of the influence of the superconductor on the QH state.

\subsection{Tight-Binding Model of the Interface \label{subsec:TB-Model}}

Consider an interface with normal-state hamiltonian
\begin{equation}
H(x,y)=H_{S}+H_{QH}+H_{T},\label{eq:Ham-full}
\end{equation}
where $H_{S}$ represents the superconductor, $H_{QH}$ the QH graphene, and $H_{T}$ the connection between them. The superconductor is described by a tight-binding model on a square lattice in the $y>0$ half-plane, and graphene in the QH state is modeled by nearest-neighbor hopping on a honeycomb lattice ($y\le0$) with phase factors due to the magnetic field. Details are given in Appendix\,\ref{app:ModelSpecifics}.  

\begin{figure}
\includegraphics[width=\linewidth]{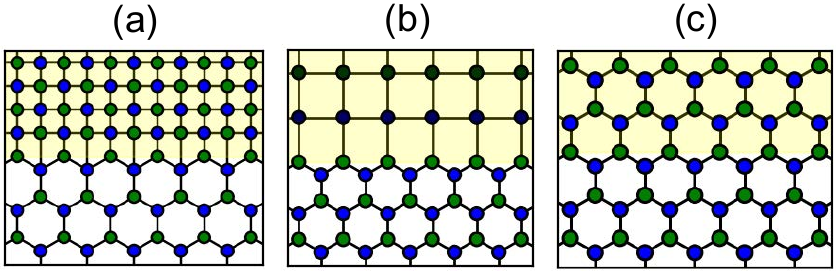}
\caption{The three 
graphene-superconductor interfaces studied with the tight-binding model: zigzag edge of graphene stitched to (a)~dense, (b)~sparse, and (c)~honeycomb lattices. The graphene sublattices---green (A) and blue (B)---extend into the superconductor for dense and honeycomb stitching. 
Atoms on the edge of the graphene sheet are in the A sublattice.
\label{fig:lattices}}
\end{figure}

Since a zigzag edge is thought to be a good model for a generic graphene edge \cite{AkhBeenBoundaryPRB08,vanOstaayReconstructPRB11,Manesco-QHgrS_SP22}, we consider a graphene half-sheet with a zigzag edge (Fig.\,\ref{fig:lattices}) \cite{ArmResults}. 
The  graphene half-sheet is translationally invariant in the $x$ direction with period $a$, the lattice constant of graphene (next-nearest-neighbor distance). Our convention is that the sites on the terminal row of the zigzag edge are at $y=0$ and belong to the A sublattice. 
For the magnetic field, we work in the Landau gauge with $\mathbf{A}(x,y)=By\hat{x}\,\theta(-y)$ where $\theta(y)$ is  the step function 
[thus $\mathbf{B}=-B\hat{z}\,\theta(-y)$]. With these choices, $\Delta$ is constant and real.

The two materials are joined by matrix elements of magnitude $t_{NS}$ that connect the graphene and superconductor edge sites. We consider three ways of ``stitching'' the two lattices together, shown in Fig.\,\ref{fig:lattices}. First, in the ``dense'' stitching scenario [Fig.\,\ref{fig:lattices}(a)], the square lattice of the superconductor has half the period of the graphene lattice and is connected to both graphene sublattices. Second, for ``sparse'' stitching [Fig.\ \ref{fig:lattices}(b)], the square lattice has the same period as the zigzag edge and only the terminal A sites are connected to the square lattice \citep{BlanterMartin_M-GrContacts_PRB07, LingfeiGlebCAESNatPhys20, TakagakiGrRibbons_JPCM21}. Third, in the ``honeycomb'' scenario, the graphene lattice is simply continued into the superconductor [Fig.\,\ref{fig:lattices}(c)], though $\mu_S$ and $\mu_\textrm{gr}$ may differ. 

The density of states in the superconductor is largest for the dense scenario. Since the period of the square lattice is half that of the zigzag edge, the band structure of the square lattice should be folded back into the smaller graphene (projected) Brillouin zone. Thus the number of bands in the dense case is approximately twice that in the sparse scenario. 

\subsection{Continuum Model of the Interface \label{subsec:ContinuumModel}}

In our continuum 
model, we match a zigzag edge of graphene
described by a Dirac hamiltonian to a superconductor with parabolic
dispersion. The matching conditions follow from wavefunction
and current continuity. The hamiltonian is again given by Eq.\,(\ref{eq:Ham-full})
where now $H_{T}$ denotes a $\delta$-function barrier separating
the two materials, 
\begin{equation}
H_{T}=V_{0}\,\delta(y).\label{eq:def_V0}
\end{equation}

First, for the \emph{superconductor}, we take a 2D free electron model
and a constant pairing potential,
\begin{align}
H_{S}(x,y>0) & =-\frac{\hbar^{2}}{2m}\nabla^{2}-\mu_{S}\label{eq:HS-contin}\\
\Delta(x,y>0) & =\Delta,\label{eq:Delta_cont}
\end{align}
where the chemical potential $\mu_{S}$ is the Fermi energy and $\Delta$ is real. For states within the superconducting gap,
$E<\Delta$, we take plane waves with wavevector $p$ along the interface
and evanescent waves perpendicular to the interface, $\psi^\beta_{S,p}(y)$ ($\beta\!=\!e$ or $h$), described by a
complex wavevector $q=q'+iq''$; see App.\,\ref{app:ModelSpecifics} for the exact form. 
This wavevector is conveniently written in terms of 
\begin{equation}
q_{y}^{2}\equiv k_{FS}^{2}-p^{2}\quad\textrm{and}\quad q_{\Delta}^{2}\equiv\frac{2m\Delta}{\hbar^{2}} , \label{eq:Def-qy-qDel}
\end{equation}
where $k_{FS}$ is the Fermi wavevector of the superconductor. 
We are particularly interested in the states at the chemical potential, $E=0$, for which we find 
\begin{align}
q^{2} & =q_{y}^{2}+iq_{\Delta}^{2},\label{eq:q_Eeq0}\\
q=q'+iq'' & \approx q_{y}+i\frac{q_{\Delta}^{2}}{2q_{y}},\label{eq:q_small-Delta-approx}
\end{align}
where the approximation for the real and imaginary parts of $q$ in
the second line can often be used because $\Delta$ is usually very
small compared to the Fermi energy. The exponential decay length in the superconductor is $\xi_{y}\equiv1/q''\approx2q_{y}/q_{\Delta}^{2}=\hbar v_{y}/\Delta$,
which we call the coherence length perpendicular to the interface.

Second, for \emph{graphene}, the components of the wavefunction are labeled by sublattice (A or B), valley ($K$ or $K'$), and electron or hole (denoted $e$ or $h$)---$\psi_{K\textrm{A}}^{e}(x,y)$, for example. 
The hamiltonian in the $K$ valley is
\begin{align}
H_{QH}(x,y<0) & = v\boldsymbol{\sigma}\cdot(-i\hbar\nabla+e\mathbf{A})-\mu_{\textrm{gr}}\sigma_{0},\label{eq:DeqElec}
\end{align}
where the Pauli matrices $\sigma_{i}$ act on the sublattice pseudospin (with $\sigma_{0}$  the unit matrix). The corresponding expression for the $K'$ valley is obtained by the transformation operator  $-i\sigma_y$.  
Using the Landau gauge, $\mathbf{A}=By\hat{x}$, we look for a plane wave solution proportional to $e^{ikx}$ where $k$ is the deviation of the particle's total wavevector from the Dirac point $K$ or $K'$. 
Solutions perpendicular to the interface, denoted e.g.\ $\psi_{K\textrm{A},k}^{e}(y)$, are written in terms of either plane waves ($B=0$) or parabolic cylinder functions (QH regime \citep{HoppeZulickePRL00,AkhmerovValleyPolarPRL07,Romanovsky_DiracQH_PRB11}), see Appendix \ref{app:ModelSpecifics}.   

The superconducting and quantum Hall forms of the wavefunction must 
match at the interface. First, because of translational symmetry along 
$x$, the total wavevector must be equal,
\begin{equation}
p=k+K\quad\textrm{or}\quad p=k+K'.
\label{eq:k-match}
\end{equation}
As these two conditions cannot be satisfied simultaneously, the amplitudes 
for valleys $K$ and $K'$ are independent.

Perpendicular to the interface, there are two boundary conditions: both the wavefunction and the current must be continuous for each type of quasiparticle (electron or hole). Since the graphene zigzag edge atoms are on the A sublattice, wavefunction continuity yields 
\begin{align} 
\psi_{\cdot\textrm{A},k}^{\beta}(y=0^{-}) & =\psi_{S,p}^{\beta}(y=0^{+}),\quad\beta=e\textrm{ or }h,\label{eq:bndcond1}
\end{align}
valid in both the $K$ and $K'$ valleys [$p$ and $k$ satisfy (\ref{eq:k-match})]. 

Current continuity constrains the velocity perpendicular to the interface and normally results in continuity of the wavefunction derivative. However, the current operator on the honeycomb lattice \citep{AkhmerovValleyPolarPRL07} relates the amplitudes on the A and B sublattices (nearest neighbors on the honeycomb lattice) and is not a derivative of the A sublattice amplitude. 
The argument in App.\,\ref{sec:BoundaryConds} shows that current continuity yields a boundary condition on the B sublattice:  
\begin{align}
v\psi_{\cdot\textrm{B},k}^{\beta}(y=0^{-}) & =\pm\frac{1}{2m}\frac{\hbar}{i}\!\left.\frac{\partial\psi_{S,p}^{\beta}}{\partial y}\right|_{y=0^{+}},\quad\beta=e\textrm{ or }h,\label{eq:bndcond2}
\end{align}
where $+$ is for the $K$ valley and $-$ for $K'$. When there is a barrier present, $H_{T}=V_{0}\delta(y)$,  it can be incorporated in the boundary conditions by a simple shift in the transverse velocity operator in Eq.\,(\ref{eq:bndcond2})
\footnote{This can be shown by placing the barrier inside the superconductor at an arbitrary position $y_{0}$, integrating across the $\delta$-function as usual for parabolic dispersion, and then taking the limit $y_{0}\to0$ \cite{BTK-PRB82,BlonderTinkhamPRB83}.},
\begin{equation}
\pm\frac{1}{2m}\frac{\hbar}{i}\frac{\partial}{\partial y}\longrightarrow\pm\frac{1}{2m}\frac{\hbar}{i}\frac{\partial}{\partial y}+i\frac{V_{0}}{\hbar}.\label{eq:V0-bndcond}
\end{equation}
Mathematically, the potential barrier increases the imaginary part of the ``momentum'' along 
the $y$ direction, increasing the decay of the wave function across the barrier. 

We emphasize that current continuity is a condition on the \textit{velocity} at the interface in the two materials. 
The relative magnitude of graphene's Dirac point velocity to the perpendicular velocity in the superconductor is clearly important in Eq.\,(\ref{eq:bndcond2}). Their ratio is a key parameter below, and so we define 
\begin{align} 
\widetilde{v}^{2}\equiv\left(\frac{v_{y}}{2v}\right)^{2} & =\left(\frac{\hbar q_{y}}{2mv}\right)^{2}.\label{eq:def-vtilde}
\end{align}
In terms of the wavevector $k_0$ of the $E=0$ state, 
\begin{align}
\widetilde{v}^{2} & =\frac{1}{4v^{2}}\frac{\hbar^{2}\left[k_{FS}^{2}-\left(K+k_0\right)^{2}\right]}{m^{2}}\approx\frac{\hbar^{2}\left(k_{FS}^{2}-K^{2}\right)}{4v^{2}m^{2}} ,\label{eq:vtildesq1}
\end{align}
where the last approximate expression follows because $\mu_\textrm{gr}$
is not far from the Dirac point and so $k_0$ is small. In order to
match the wavevector along the interface, Eq.\,(\ref{eq:k-match}),
$\widetilde{v}^{2}>0$ is required in order that $\mu_S >\hbar^{2}K^{2}/2m$.

Using these boundary conditions in the lowest Landau level spin-degenerate
case, we find one solution in each valley, $E_{K}(k)$ and $E_{K'}(k)$ 
(see App.\,\ref{app:ModelSpecifics} for details).

\subsection{Regime Studied and Parameter Values \label{subsec:Parameters}}

 Following the experiments \citep{LingfeiGlebCAESNatPhys20,LingfeiGlebLossPRL23,SahuDas_Noise_PRB21,HatefipourShabani_QH-S_NL22}, 
we assume a hierarchy of energy scales. Superconductivity is characterized by the smallest scale, $\Delta$, and the largest scale is the bandwidth or Fermi energy in the superconductor. The magnetic energy is intermediate: we characterize it by the energy difference between the $n=0$ and $n=1$ Landau levels, denoted $E_{L}\equiv\sqrt{2}\hbar v/l_{B}$. In addition, the density of conduction electrons in graphene is much lower than in the superconductor. Thus we consider the regime  
\begin{equation} 
\Delta\ll E_L,\mu_{\textrm{gr}}\ll \mu_S. 
\end{equation} 
With regard to length scales, the magnetic length ($l_{B}^{2}=\hbar/eB$) should be much larger than the lattice constant in either material. We use a standard set of parameter values for most of the results presented and note any deviations from these standard values. 

In the tight-binding model, the hopping matrix element in both the graphene and superconductor is $t$.
Our standard parameter values are $t_{NS}\!=\!t$, $E_{L}/t\!=\!0.12$, $\mu_\textrm{gr}/E_{L}\! =\!1/4$, and $\Delta/E_{L}\!=\!1/10$  
\cite{WidthFootnote}. 
The magnetic length and magnetic field corresponding to this $E_{L}$ are $l_{B}/a\!=\!10.2$ and $Ba^{2}/\Phi_{0}\!=\!0.0015$ (or a flux of $0.0013\,\Phi_{0}$ per unit cell), where $a$ is the lattice constant of graphene. In the superconductor, $\mu_{S}$ is typically around half filling, yielding $\mu_{S}/t \!\sim\! 4$. 
(The ``optimal values'' of $\mu_S/t$ that we use are $4.5$, $5.0$, and $0.2$ for the dense, sparse, and honeycomb lattices, respectively.) Thus the Fermi energy in graphene is much smaller than that in the superconductor: $\mu_\textrm{gr}/\mu_{S}\sim0.01$.

In the continuum model, the most important parameter is the velocity
ratio (\ref{eq:def-vtilde}) entering the current continuity boundary
condition: our standard value is $\widetilde{v}^{2}=1$, which corresponds
to perfect matching. The strength of the superconductivity is set
by $\Delta=0.01(2mv^{2})$, ensuring slow evanescent decay in the
superconductor, $\xi_{y}q_{y}=200$ in the approximation of Eq.\,(\ref{eq:q_small-Delta-approx}).
The magnetic field is fixed by either $\xi_{y}/l_{B}\!=\!10$ or $\Delta/E_{L}\!=\!0.1$,
which differ by only a factor of $\sqrt{2}$. We use quarter filling
of the LLL edge state as our standard doping, $\mu_\textrm{gr}/E_{L}\!=\!1/4$, 
and no graphene-superconductor barrier, $V_{0}=0$.

In the absence of spin-orbit coupling and the Zeeman effect,  the problem decouples into
two independent, identical problems, leading to doubly
degenerate states. We therefore neglect spin until Zeeman effects are discussed in Sec.\,\ref{sec:Spin-Zeeman}.

\section{Strong Andreev Reflection \label{sec:StrongAndreev}}

We start by demonstrating that it is possible to get strong Andreev
reflection at a graphene-superconductor interface, both at $B=0$
and in the quantum Hall regime, despite the abrupt change in material
and the very large mismatch in the density of states.  This stands
in sharp contrast to the well-established result for semiconductor-superconductor
interfaces (for parabolic dispersion in both) 
in which the probability of Andreev reflection is small (except for special fine-tuned cases) \citep{BTK-PRB82, BlonderTinkhamPRB83, NakanoBTKinterferomPRB93, MortensenAngleAndreevPRB99, TakagakiQHEPRB98, GiazottoPRB05, ADavid_Grenoble23}.
For clarity, we focus on only a few values of parameters in this section
and postpone (Secs.\,\ref{sec:Dispersion} and \ref{sec:Vary-E_Fsup})
a discussion of the sensitivity to model parameters.

\subsection{$B=0$: Robust Andreev Reflection and the Dirac Point Velocity} \label{subsec:Zero-Field}

Before turning to the QH regime, it is instructive to consider the
$B=0$ case. An electron incident on the interface  will be reflected
as either an electron or hole. The probability of Andreev reflection
follows from wavefunction matching. This was done for an interface
between free-electron materials (parabolic dispersion) in the classic
work by Blonder, Tinkham, and Klapwjik (BTK)  \citep{BTK-PRB82,BlonderTinkhamPRB83}.
Here we perform a similar analysis for an interface between graphene
described by the continuum Dirac equation, Eq.\,(\ref{eq:DeqElec}) with
$\mathbf{A}=0$, and a superconductor with parabolic dispersion, Eq.\,(\ref{eq:HS-contin}). 

The form of the wavefunction in the superconductor does not depend
on $B$ and so is the same as in the previous section. On the graphene
side, we continue to look for a 
solution proportional to
$e^{ikx}$ but now the solution should be plane wave perpendicular
to the interface as well, with an ingoing electron state and outgoing
electron and hole states. The calculation is presented
in App.\,\ref{sec:B=0Calc}, together with additional results.

The main feature is already evident in the simple case of normal incidence
in the limit $q_{\Delta}\ll q_{y}$. In this limit, Eq.\,(\ref{eq:q_small-Delta-approx})
implies that the imaginary part of $q$ can be neglected in the boundary
conditions. We find in this case  that the probability of Andreev reflection is 
\begin{align}
P_{A} & =\frac{4\widetilde{v}^{2}}{\left(1+\widetilde{v}^{2}\right)^{2}},\label{eq:PA_B0-normal}
\end{align}
where $\tilde{v}$ is the ratio of velocities defined in Eqs.\,(\ref{eq:def-vtilde})-(\ref{eq:vtildesq1}).
The functional dependence here is striking: there is a broad maximum
of perfect Andreev reflection at $\widetilde{v}^{2}=1$, while $P_{A}\to0$
in the two limits $\widetilde{v}^{2}\to0$ and $\widetilde{v}^{2}\to\infty$.
The characteristics of the superconductor enter through a somewhat
weak dependence on the transverse velocity $v_{y}$---note that $k_{FS}$
and $K$ do not enter independently. The key point, however, is that
the dependence on $k_0$, and thus on $\mu_\textrm{gr}$, in (\ref{eq:vtildesq1}) is very weak.
The wavelength of the particles in graphene plays almost no role. 
\emph{Andreev reflection in clean graphene is determined by the properties of the
superconductor and the conical valley structure of graphene---the
value of $K$ and, most importantly, the velocity at the Dirac point.}

Quantitatively, the velocity at the Dirac point in graphene is of
the same order as the Fermi velocity in metals, so from (\ref{eq:PA_B0-normal})
we can expect situations in which there is very good Andreev reflection,
\begin{align}
\widetilde{v}\sim1 & \quad\textrm{and}\quad P_{A}\lesssim1,
\end{align}
even for a large mismatch between the graphene and the superconductor,
$\mu_\textrm{gr} \ll \mu_{FS}$. While these results are for the simplest
case of normal incidence neglecting the effect of $q_{\Delta}$, results
in App.\,\ref{sec:B=0Calc} go beyond those assumptions and
show that the conclusion is general. 

This is very different from Andreev reflection in a semiconductor
with parabolic dispersion. In that case, the big mismatch between
the velocities in the semiconductor and the superconductor \cite{v-parabolic} suppresses
Andreev reflection strongly. With the same simplifications
as for graphene above, the result for a clean semiconductor-superconductor
interface is
\begin{align}
P_{A} & \approx\dfrac{4}{\widetilde{v}^{2}}=4\frac{m_{S}\mu_{FN}}{m_{N}\mu_{FS}}\ll1,
\label{eq:PA-B0semicond}
\end{align}
where $N$ denotes the semiconductor \citep{BTK-PRB82, BlonderTinkhamPRB83, NakanoBTKinterferomPRB93, MortensenAngleAndreevPRB99}.
\emph{Thus, the conical valleys in graphene allow much better connection
to superconductors than for semiconductors with comparable density: the honeycomb lattice provides a natural decoupling of the Fermi velocity (appearing in current continuity) from the Fermi wavevector.}  
As a result, in graphene there is the possibility of near perfect Andreev
reflection. 

\subsection{Quantum Hall Regime: Good Electron-Hole Hybridization \label{subsec:Good-eh-Hybrid}
}

In this subsection, we show that good electron-hole hybridization in the QH regime
is possible.  By ``good hybridization'' we mean
that the weight of electrons and holes in each CAEM wavefunction are nearly equal [$f_e \approx 0.5$, see Eq.\,(\ref{eq:fe-define})]. 

\subsubsection{QH Edge States on a Terminated Lattice}

\begin{figure}
\includegraphics[width=3.3in]{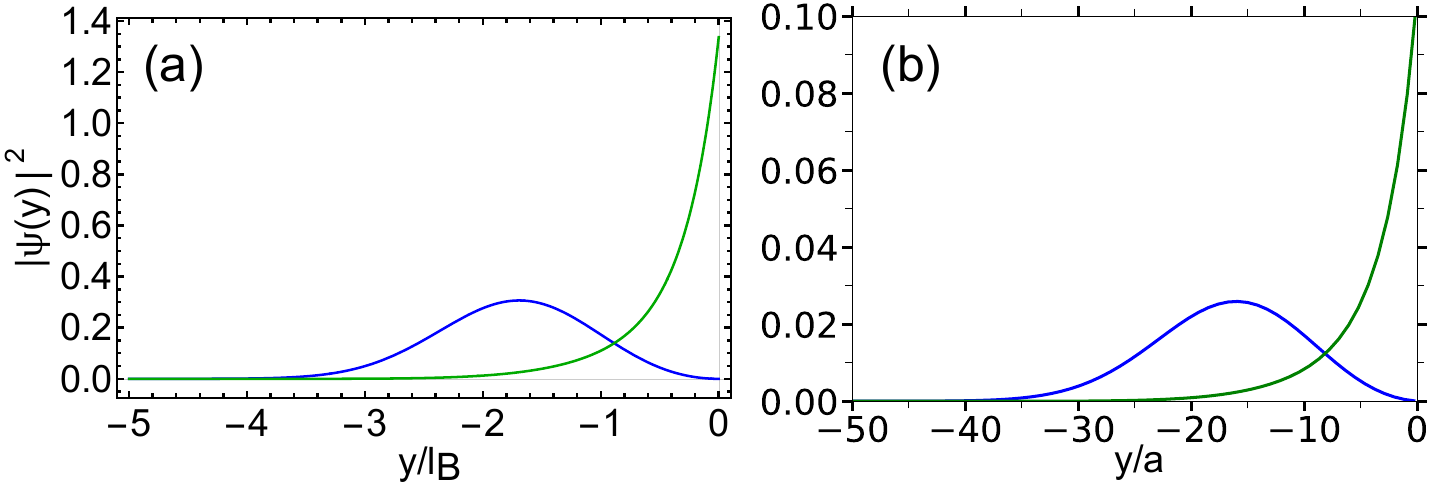}
\caption{QH edge states of a terminated graphene
lattice (zigzag edge) in the (a) continuum and (b) tight-binding models. The wavefunction
of the $K$-valley mode, $|\psi_K(y)|^{2}$, on the A (green) and
B (blue) sublattices is shown. (Standard parameters, Sec.\,\ref{subsec:Parameters}.)}
\label{fig:termin-latt-psi}
\end{figure}

We start with the QH states on the graphene edge
in the absence of a superconductor---that is, when the lattice is
abruptly terminated. The electron states in this case are well-known
for both the tight-binding and continuum models \citep{BreyFertig_EdgePRB06,CastroNeto_GrEdge_PRB06,AkhmerovValleyPolarPRL07,DelplaceMontambauxPRB10,Romanovsky_DiracQH_PRB11,CastroNeto_GrRev_RMP09}. 

In the bulk, the LLL states have energy zero (with respect to the
Dirac point). In the Landau gauge used here, the wavefunctions are
plane waves in $x$ and a simple Gaussian in $y$ around a guiding
center $\overline{y}(k)$. For each guiding center position, there are two solutions
corresponding to measuring $k$ from either $K$ or $K'$. Importantly,
there is \emph{``sublattice-valley locking''}: solutions associated
with valley $K$ are on the B sublattice exclusively
and those associated with $K'$ are on the A sublattice.

When $\overline{y}$ approaches the $y\!=\!0$ edge, boundary conditions come into play.  
In our geometry the boundary condition is
that $\psi$ on the B sublattice must be zero (i.e.\,the absent
next row in the lattice). Since the $K'$ valley states are non-zero
only on A sites, they are not modified: their energy remains strictly
zero.

In contrast, 
the $K$ valley wavefunctions reside on the B sites and must
go to zero faster at the edge than in the bulk, changing their energy.
Thus, the LLL edge mode at the chemical potential corresponds to an
electron in the $K$ valley. With the change in $\psi$ on B sites,
there must be amplitude on A sites in order to satisfy the coupled
equations. The wavefunction is shown for both the continuum and tight-binding
models in Fig.\,\ref{fig:termin-latt-psi}; results of the two approaches are in very good agreement. 

\begin{figure}[b]
\includegraphics[width=2.5in]{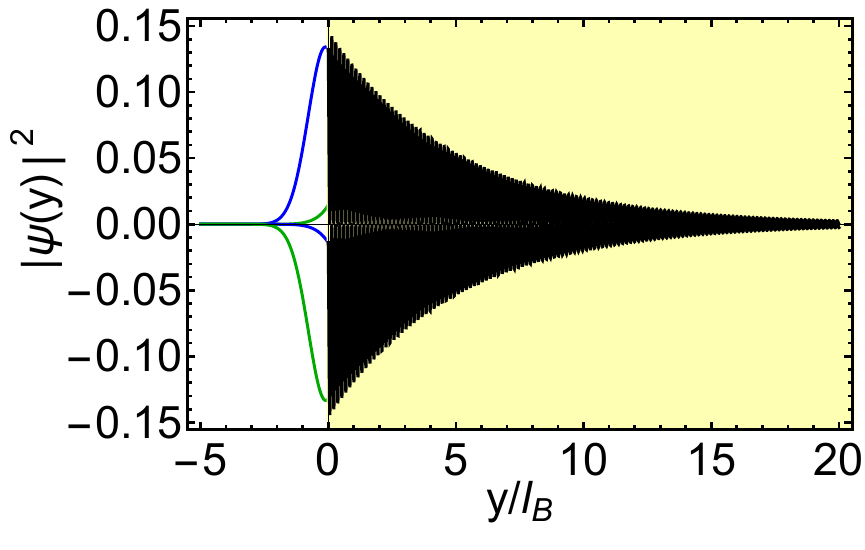}
\caption{Wavefunction of the interface mode, $|\psi(y)|^{2}$,
in the continuum model for ideal graphene-superconductor matching:  for $E\!=\!0$ in the $K$ valley 
as in Fig.\,\ref{fig:MainResults}(a), but on an expanded scale to show the evanescent decay
into the superconductor (Standard parameters, Sec.\,\ref{subsec:Parameters}.)}
\label{fig:psi-continuum}
\end{figure}

\begin{figure*}
\includegraphics[width=\textwidth]{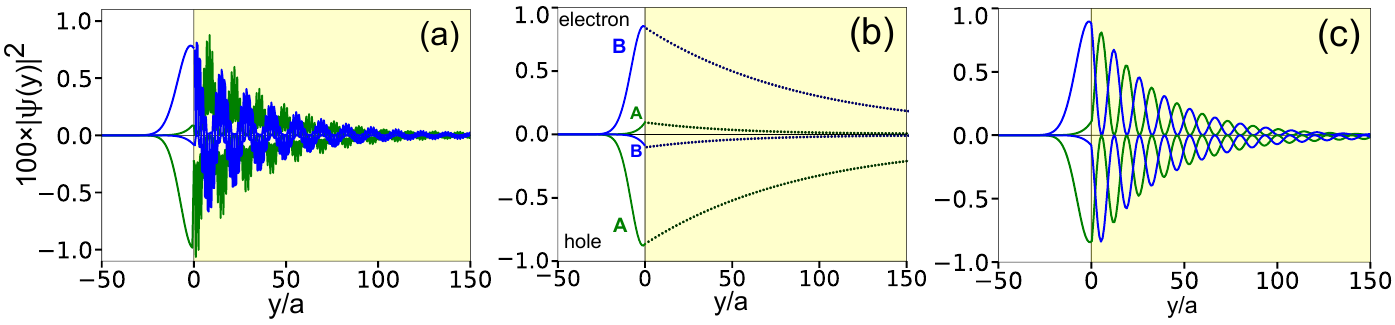} 
\caption{Mode wavefunctions, $|\psi(y)|^{2}$,
at the chemical potential ($E=0$) in the $K$-valley for three
well-matched tight-binding interfaces: (a) dense, (b) sparse, and (c) honeycomb superconductor (shown in Fig.\,\ref{fig:lattices}). The electron and hole
components are plotted as \emph{positive} and \emph{negative} numbers, respectively,
and green (blue) refers to the A (B) sublattice. All three
interfaces are highly transparent, achieved using $t_{NS}/t=1$ and $\mu_{S}/t=$ (a)
$3.8$, (b) $5$, and (c) $0.2$. In panel (b) for $y>0$, dark green and
blue denote the even and odd rows of the square lattice
counting from the interface. 
The degree of $e$-$h$ hybridization is given in Table \ref{tab:T-fe-dk_forFigs}. (Standard parameters, Sec.\,\ref{subsec:Parameters}.)\label{fig:psi-tightbinding}}
\end{figure*}

\subsubsection{Interface wave function: Continuum model \label{subsec:ehHyb-Continuum}}

For the continuum model at $B=0$, we found perfect Andreev reflection
if the superconductor and graphene are well matched, $\widetilde{v}^{2}\approx1$.
The corresponding wavefunction in the QH regime is shown in Figs.\,\ref{fig:MainResults}(a)
and \ref{fig:psi-continuum} for $\widetilde{v}^{2}=1$. 
Strikingly, the magnitude and spatial dependence of the electron component in graphene ($y<0$) are 
the same as those of the holes, except that the sublattice components are interchanged. 
Evidently in this case, Andreev reflection in the QH regime is also perfect. 

The wavefunction in graphene is clearly very different from that for
a terminated lattice (Fig.\,\ref{fig:termin-latt-psi}). Here, the
electron resides mostly on one sublattice (B, blue)---that of
the bulk, $K$-valley LLL wavefunction---and is gaussian shaped,
centered slightly left of the interface, and cut at $y=0$. The deviation
from gaussianity produces the small amplitude exponential decay on
sublattice A.

In the hole component, the roles of sublattices A and B are reversed. 
Because the sublattice structure is different, 
the hole component here \emph{cannot} represent 
the absence of the same type of electron as in the electron component, namely a $K$-valley electron. Rather, the hole component
is similar to the bulk LLL wavefunction for the $K'$ valley which
is on the  A sublattice (sublattice-valley locking). Thus, a $K$-valley electron has been paired 
with a $K'$ valley electron, as expected for a Cooper pair. 

Three length scales are visible in the wavefunction. (i)~The
magnetic length, $l_{B}$, sets the scale of variation in the quantum
Hall state. (ii)~The Fermi wavelength in the superconductor sets
the scale for the rapid oscillations at $y>0$. (iii)~The scale
of the decaying envelope in the superconductor is 
the coherence length $\propto1/\Delta$ 
\footnote{More precisely, for quantities (ii) and (iii) one must take the projection
perpendicular to the interface [see Eq.\,(\ref{eq:q_Eeq0})]:
for (ii), the period of the oscillations is $2\pi/q' \!\approx\! 2\pi/q_{y}$,
and for (iii), the coherence length perpendicular to the interface
is $\xi_{y}\equiv1/q'' \!\approx\! \hbar v_{y}/\Delta$ which, for small
$\Delta$, is much larger than the period. The rapid oscillations
of the electron and hole wavefunctions in the superconductor are exactly
out of phase, as expected for a mid-gap state [$E=0$ implies $\gamma=-i$ in Eq.\,(\ref{eq:psiS})].}.

The filling of the edge state, $\mu_\textrm{gr}/E_L$,
plays little  role in the matching. Indeed, for a suitable superconductor for which $\widetilde{v}^{2}\approx 1$, the degree of electron-hole hybridization is essentially independent of $\mu_{\textrm{gr}}$.  The fractional electron content in graphene, $f_e$ in Eq.\,(\ref{eq:fe-define}), is one way to quantify the $e$-$h$ hybridization. 
Fig.\,\ref{fig:MainResults}(b) shows $f_{e}$ in the LLL as $\mu_\textrm{gr}$ varies from the Dirac point to the threshold for the first Landau level, $E_{L}$. 
For $\widetilde{v}^{2}=1$, hybridization is perfect, $f_{e}=1/2$, throughout the LLL.  Thus, as in the $B\!=\!0$ case, we see that the decoupling of the Fermi velocity and conduction electron density enables strong Andreev reflection.

\subsubsection{Interface wave function: Tight-binding model \label{subsec:ehHyb-Tightbind}}

\begin{table}
\begin{tabular}{c c c}
\parbox[t]{0.5in}{Fig. (panel)} & \parbox[t]{0.72in}{Fractional electron content, $f_{e}$} & 
\parbox[t]{0.98in}{Deviation from valley center, $\delta k_0\,l_{B}$} \\
\hline 
3(b) & 1.0 & 1.49 \\
5(a) & 0.50 & 0.10 \\
5(b) & 0.49 & 0.02\\
5(c) & 0.54 & 0.05\\
12(a)  & 0.82 & 0.40\\
12(b)  & 0.78 & 0.37\\
\hline
\end{tabular}
\smallskip
\caption{\label{tab:T-fe-dk_forFigs}For the tight-binding wave functions shown
in Figs.\,\ref{fig:termin-latt-psi}, \ref{fig:psi-tightbinding}, and \ref{fig:psi_largeFsurf},
we give the fractional electron content in graphene [$e$-$h$ hybridization, Eq.\,(\ref{eq:fe-define})] and the deviation of the wavevector $k_0$ for the $E=0$ mode  from the center of the $K$ valley: $\delta k_0\,l_{B}\equiv(k_0-K)l_{B}$.}
\end{table}

Our tight-binding results corroborate the features found in the continuum limit. 
The wavefunctions for the three superconductor models are shown in Fig.\,\ref{fig:psi-tightbinding}. 
In each case we have chosen the Fermi energy in the superconductor to achieve very good matching to the graphene, and the transparency of the interface is about 85\%.  (By ``transparency'' here we mean specifically the transmission probability at $B\!=\!0$ in the normal state \cite{Transparency-def}.) 
Note that there is nearly equal weight of electrons and holes in graphene. Indeed, on both sides of the interface and in all three cases, the hole component is directly related to the electron component by simply interchanging the A and B sublattices. 

It is clear from the plots that there is excellent global electron-hole hybridization. Quantitatively, 
the value of $f_e$ is found from the wavefunction [see Eqs.\,(\ref{eq:fe-define}) and (\ref{eq:fe-def-tbind})] and reported in Table \ref{tab:T-fe-dk_forFigs}. They are $\approx\! 0.5$ and so nearly ideal. However, we wish to emphasize a more striking feature: if we consider the sum of the A and B site weights, the spatial profile of the electron and hole components is nearly the same.  Thus, there is excellent electron-hole hybridization in every graphene unit cell. 

On the graphene QH side ($y<0$), the wavefunction for each stitching has all the characteristics of the continuum. The B sublattice predominates in the electron component, corresponding to the bulk LLL state in the $K$ valley, and the shape of the B-sublattice component is close to that of a bulk wavefunction. To be consistent, $\psi(y)$ on the A sublattice is small, in contrast to the QH edge state [Fig.\,\ref{fig:termin-latt-psi}(b)].  
In the hole component, the two sublattices are interchanged. The shapes of the electron and hole components are very similar but not identical---the fact that the QH side terminates in A sites breaks the symmetry. 

On the superconductor side ($y>0$), the three cases are similar despite the different underlying lattices. First, the exponential decay of the envelope of oscillations is controlled by the magnitude of the superconducting gap, $\xi_{y}\approx\hbar v_{y}/\Delta$, as in the continuum model. In the honeycomb-lattice superconductor, panel (c), the sublattice structure extends into the superconductor, and oscillations corresponding to the transverse wavevector occur with a period shorter than $l_{B}$ because $\mu_{S}>\mu_{\textrm{gr}}$. The fact that the oscillations on the A and B sublattice are out of phase is connected to the big difference in weight of the two sublattices on the QH side.  

The square lattice is bipartite, of course, and for dense-stitching, the bipartite nature on the two sides of the interface match [see Fig.\,\ref{fig:lattices}(a)]. Thus, panel (a) shows oscillating sublattice weights in the superconductor, much as in (c). The very fast oscillation is the wavelength in the superconductor: since $\mu_{S}$ is near half filling, $\lambda_{FS}$ is on the lattice scale.  

With sparse stitching, there is only one square lattice site per graphene unit cell and so once stitched together the lattice is not bipartite. Focusing on a unit cell of the interface, however, we see that the square lattice consists of a single chain of sites, and so the bipartite structure is reduced to even-odd alternation. Indeed, in panel (b), the wavefunction oscillates rapidly between even and odd rows in the superconductor ($\mu_{S}$ is near half filling) 
\footnote{The decay of $\psi_{S}$ is slower in Fig.\,\ref{fig:psi-tightbinding}(b) 
than in (a) because the lattice constant is double that in the dense  case.}.  
It is particularly striking that the electron and hole weights for $y>0$ are exactly out of phase.

\section{Dispersion of Chiral Andreev Edge Modes: Valley Degeneracy? \label{sec:Dispersion}}

Having established the possibility of strong hybridization in the
CAEM, we now further investigate their properties through
the spectrum of states. 

\subsection{QH Terminated-Lattice Spectrum}

Before discussing the spectrum of the interface itself, we present
the BdG spectrum for an abruptly terminated honeycomb lattice in the
QH regime. The spectrum of the electron solutions is well-known \citep{BreyFertig_EdgePRB06, CastroNeto_GrEdge_PRB06, AkhmerovValleyPolarPRL07, DelplaceMontambauxPRB10, Romanovsky_DiracQH_PRB11, CastroNeto_GrRev_RMP09}.
To obtain the hole spectrum from the electrons, the particle-hole
symmetry of the BdG equation implies that one inverts both the crystal
momentum and the energy \cite{[{}][{ p.\,500.}]ColemanBook}, 
\begin{equation}
K+k\to K'-k\quad\textrm{and}\quad E\to-E.
\end{equation}

\begin{figure}
\includegraphics[width=\linewidth]{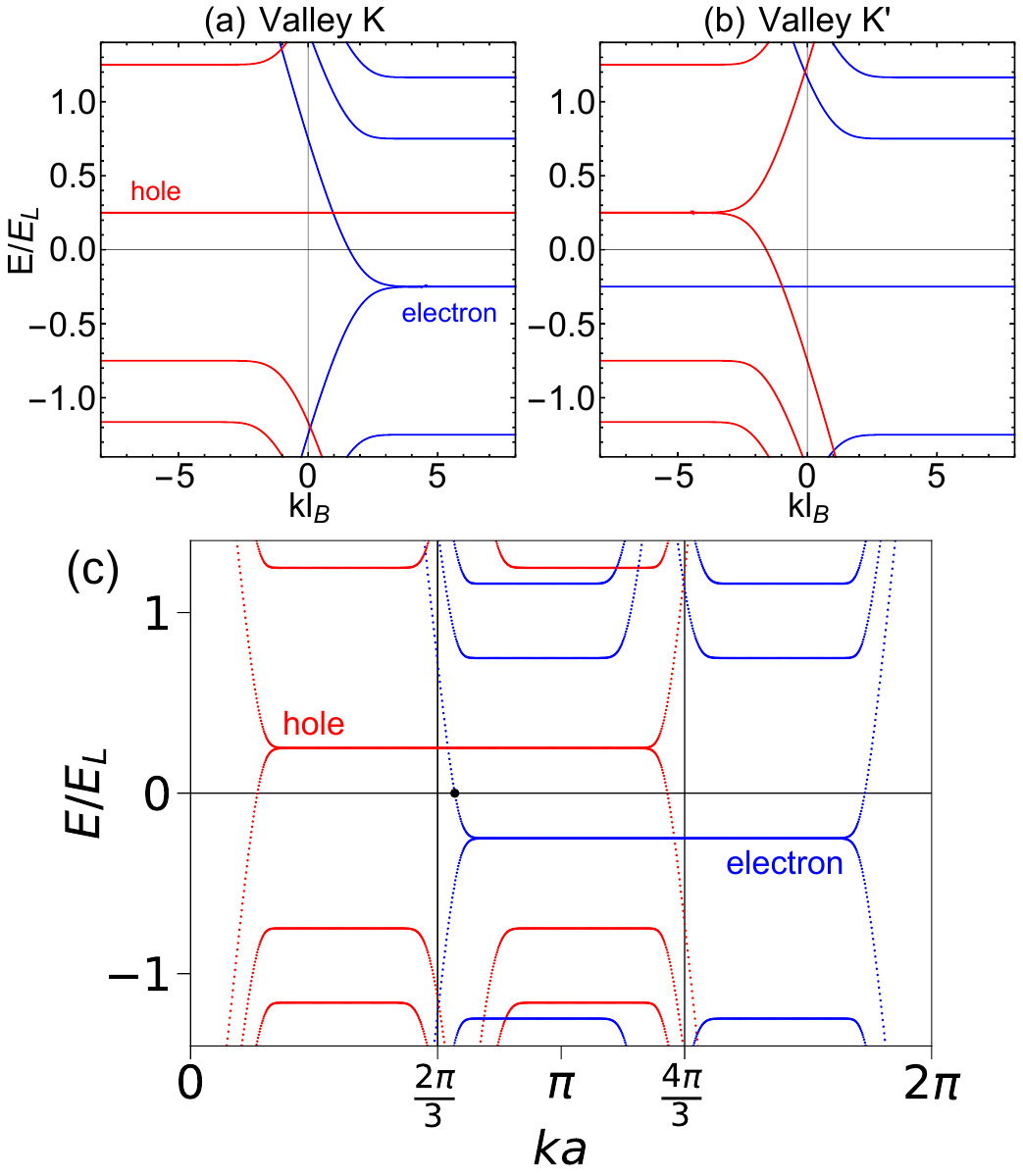}
\caption{Spectrum of electron (blue) and hole (red) states for a zigzag nanoribbon in the QH regime. (a)-(b) Continuum model for the $K$ and $K'$ valley, respectively. 
(c) Tight-binding results: the Brillouin zone is shifted to $[0,2\pi/a]$, and the hole spectrum is translated by a reciprocal lattice vector into this interval. 
The point marks the electron edge state, $E(k_0)\!=\!0$. The value of $k_0$ given in Table \ref{tab:T-fe-dk_forFigs} shows that this state is located $\sim \! l_B$ inside the graphene [wavefunction shown in Fig.\,\ref{fig:termin-latt-psi}(b)]. 
(Standard parameters, Sec.\,\ref{subsec:Parameters}.)}
\label{fig:spectrum_term-latt}
\end{figure}

The spectrum with both electrons and holes is shown in Fig.\,\ref{fig:spectrum_term-latt}
\cite{BdG-redundancy}. In the continuum solution, panels (a)-(b), both the electron mode
at the chemical potential in the $K$ valley and the corresponding
hole mode in $K'$ are clearly seen. Since the LLL $K'$ valley electron states
are not modified by the termination of the lattice (they are non-zero
only on A sites), their energy remains zero: they appear as a horizontal blue line in panel (b). Due to this flat band,
while there is no hole state in the $K$ valley at the chemical potential
(no red line crosses $E=0$), the electron and hole solutions do intersect
at $E=\mu_{\textrm{gr}}$. 

For the tight-binding results, panel (c), periodicity in the $x$
direction leads to a Brillouin zone of
size $2\pi/a$.  We show the interval $k\in[0,2\pi/a]$ in which the Dirac points are $Ka\!=\!2\pi/3$ and $K'a\!=\!4\pi/3$. To focus on the $y\!=\!0$ edge of the nanoribbon, the width of the nanoribbon, $W$, is chosen so that
the spectrum of the states on the other edge is clearly separated: $W\lesssim l_{B}^{2}(4\pi/3a)$ 
\footnote{The choice $W\!\lesssim l_{B}^{2}(4\pi/3a)$ avoids having the spectrum of the bulk states wrap around the Brillouin zone, which would then overlap the spectrum of the edge states near $y\!=\!0$.}. 
With the gauge that we use, the QH states with guiding center coordinate at the interface, $\overline{y}\!=\!0$, correspond to $k=K\!=\!2\pi/3a$ and $k=K'\!=\!4\pi/3a$.

At the chemical potential, there is an electron state in the $K$ valley 
and a hole in $K'$.
Note that the spectrum near the Dirac points 
is very similar to that of the continuum model (see also the zoom near the $K$ point in Sec.\,\ref{subsec:SecularEq}). The tight-binding solution shows 
how the two valleys are connected through the $M$ point, $ka=\pi$.

\subsection{Implications for Andreev Reflection\label{subsec:AndreevImplications} }

From these results on a terminated lattice, \emph{what can we anticipate
for the interface with a superconductor?} 

At first sight, the case for a strong proximity effect appears hopeless: we should look for electron and hole states that 
(i)~have the same crystal momentum (translational invariance), 
(ii)~overlap spatially in the transverse direction (local interaction), and 
(iii)~differ in energy by $\lesssim\Delta$ (matrix element of superconductivity
that mixes electrons and holes). 
These criteria are \emph{never} satisfied in Fig.\,\ref{fig:spectrum_term-latt}: there is no crossing of the electron and hole modes at the chemical potential. 

One clear scenario that avoids this dead end is to abandon the assumption
of a clean system \citep{Manesco-QHgrS_SP22,KurilovichDisorderAES_NCom23}. Disorder breaks translational invariance, thus
removing criterion (i) above, allowing the $E=0$ $K$-valley electron
and $K'$-valley hole in Fig.\,\ref{fig:spectrum_term-latt} to be
coupled by the superconductivity matrix element $\Delta$. The role
of disorder in the superconductor when tunneling to a clean graphene
edge has been emphasized recently in Ref.\,\citep{KurilovichDisorderAES_NCom23}. 

\begin{figure}
\includegraphics[width=3.2in]{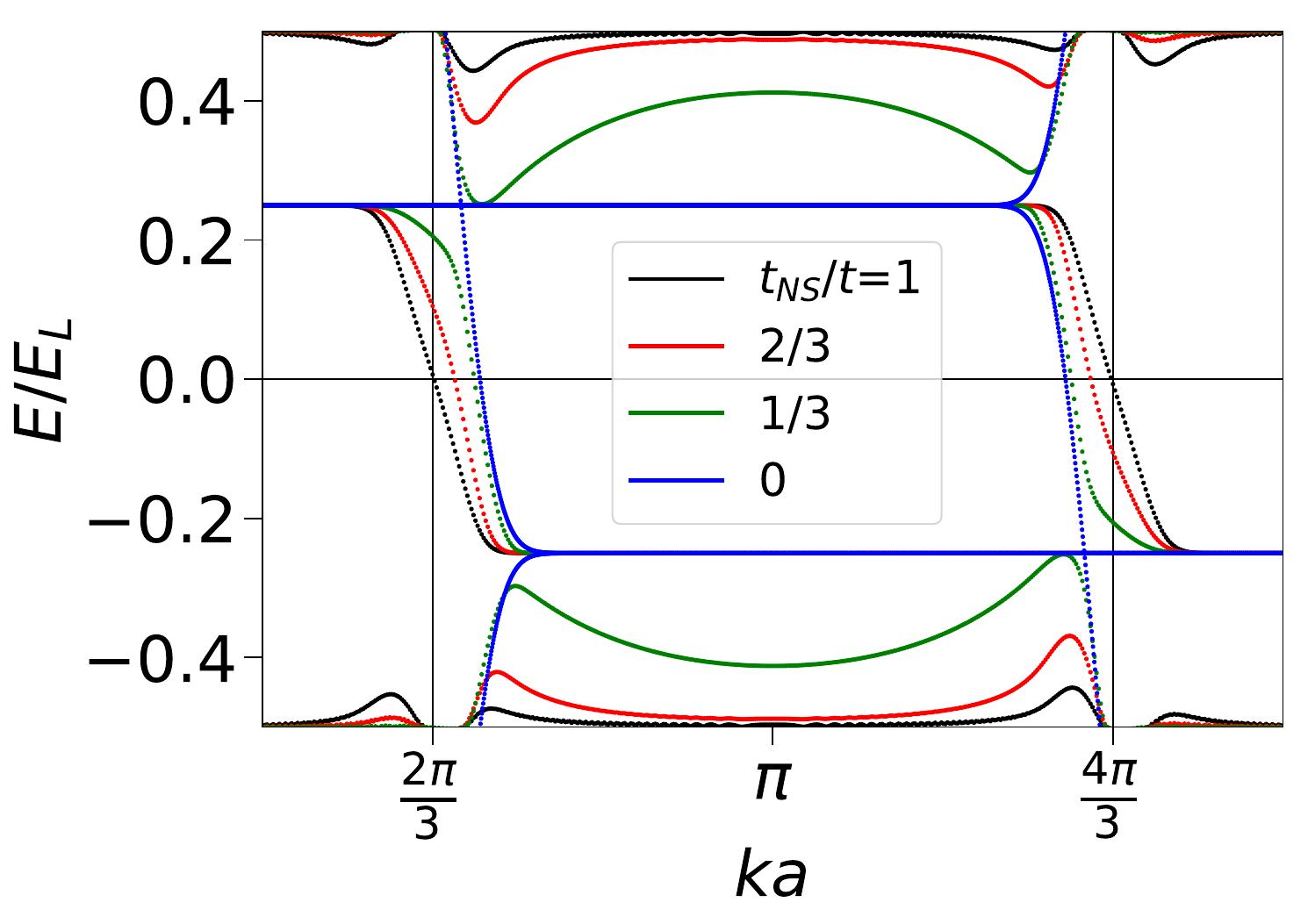}
\caption{Edge mode dispersion for different
strengths of coupling to the superconductor: tight-binding results
for sparse stitching and large pairing gap $\Delta/E_{L}=0.5$ ($\Delta>\mu_\textrm{gr}$).
The values $t_{NS}/t\in\{0,1/3,2/3,1\}$ correspond to interface transparency
$\{0,0.25,0.7,0.85\}$, respectively \cite{Transparency-def}.  The spectrum evolves from
the terminated-lattice edge mode to the fully-hybridized interface
mode centered on the $K$ point. (Otherwise standard parameters, Sec.\,\ref{subsec:Parameters}.)}
\label{fig:TBdispersion-tNS-1}
\end{figure}

However, we find that there is a second scenario that leads to a
strong proximity effect in the \emph{clean} limit, namely if the interface
is transparent. Fig.\,\ref{fig:TBdispersion-tNS-1} shows the evolution
of the spectrum as the graphene-super\-con\-duc\-tor hopping matrix element,
$t_{NS}$, increases. (In order to clearly see the LLL states at $E=\pm\mu_{\textrm{gr}}$,
a large value of $\Delta$ is used, $\Delta/\mu_{\textrm{gr}}\!=\!2$, so that they are not hidden by the superconductor's continuum.) 

In the disconnected case (blue dots), there is a crossing in each valley between three states: in the $K$ valley, for instance [see also Fig.\,\ref{fig:spectrum_term-latt}(a)], the rapidly dispersing terminated-lattice edge state intersects two non-dispersive hole states at $E\!=\!\mu_{\textrm{gr}}$ (horizontal blue line).
These holes correspond to the degenerate electron states at $E\!=\!-\mu_{\textrm{gr}}$ in the $K'$ valley, whose
nature is discussed in the graphene nanoribbon literature \cite{BreyFertig_EdgePRB06, CastroNeto_GrEdge_PRB06, DelplaceMontambauxPRB10, vanOstaayReconstructPRB11, Rodrigues_RecZagPRB11}.  
We learn that one of the non-dispersive hole states is the normal $K$-valley LLL hole state, which is not dispersive at the edge because it has weight on only the A sublattice while the boundary condition is on the B sublattice.  
The other non-dispersive hole state is necessarily far from the $y\!=\!0$ interface: depending on the nanoribbon width, it is either the $K'$ valley LLL state with guiding center deep in the bulk or the anomalous zigzag-nanoribbon state localized on the far edge of the nanoribbon.

For small $t_{NS}$ (green), an avoided crossing appears between the two states near the interface: the electron edge state and the normal $K$-valley hole state avoid crossing, while the state far from $y\!=\!0$ remains unchanged. 
The avoided crossing indicates electron-hole hybridization. 

As $t_{NS}$ grows, the spectrum
changes throughout the Brillouin zone and goes well beyond
a simple avoided crossing: one state is pushed to higher energy
while the other becomes a CAEM with antisymmetry about the $K$ point
(black). The large hopping matrix element at a transparent interface 
can couple many Landau levels, $t_{NS}\!\gg\! E_{L}$ (in Fig.\,\ref{fig:TBdispersion-tNS-1}, $t/E_L\!=\!8$), and 
reorganizes the electron and hole states, leading to electrons and holes at nearly the same energy. The small pairing matrix element $\Delta$ then readily 
causes electron-hole hybridization.  The big reorganization of the electron and hole states implies that their original non-dispersive character due to special zigzag-nanotube characteristics is not important 
\cite{NotSpecialZZ}: 
\emph{the CAEM are produced by a combination of large $t_{NS}$ and small $\Delta$ and cannot be thought of as hybridization between only two terminated-lattice states.} 

We conclude that while a superconductor cannot induce a proximity effect on a clean zigzag QH edge through 
tunneling---the dead end applies when $t_{NS}\!\ll\! t$---a strong proximity effect
can readily occur when $t_{NS}\!\sim\! t$ so that the interface is transparent.
Notably, the induced pairing correlations are quite different in these two
scenarios. In the first (disorder, tunneling), sublattice-valley locking implies that the graphene sublattice structure of the electrons and holes is the same. 
In the second scenario (clean, transparent), the electrons and holes are on opposite sublattices, as in Figs.\,\ref{fig:MainResults}(a) and \ref{fig:psi-tightbinding}. 

\subsection{Spectrum of the CAEM}

\begin{figure}
\includegraphics[width=3.07in]{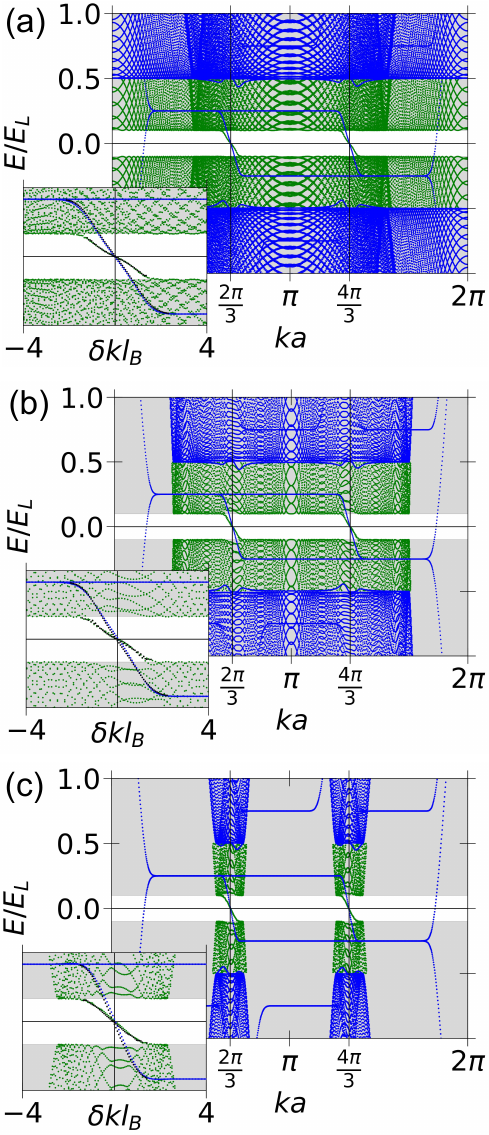}
\caption{Energy of BdG states (electrons and holes)
for interfaces with the three different superconductors---(a)\ dense
square lattice, (b)\  sparse square lattice, and (c)\ honeycomb
lattice (see Fig.\,\ref{fig:lattices})---for two values of the superconducting gap, $\Delta/E_{L}=0.1$
(green) and $0.5$ (blue). In each case, $\mu_{S}$ is chosen so that
the superconductor and graphene are well-matched: $\mu_{S}/t=4.5$,
$5$, $0.2$ for dense, sparse, and honeycomb, respectively. For
energies above the superconducting gap, the spectrum becomes dense
due to the large density of states in the superconductor. Insets:
Zoom on the region around the $K$ point in which $-E_{K}(-\delta k)$ is also plotted 
(black). The fact that the black data nearly perfectly overlay the
blue and green demonstrates excellent valley degeneracy. (Standard
parameters, Sec.\,\ref{subsec:Parameters}.)}
\label{fig:Spectra-matched}
\end{figure}

We now turn to discussing in more detail the energy spectrum of the
graphene-superconductor system.  Fig.\,\ref{fig:Spectra-matched}
shows an example for each stitching when the superconductor is well-matched
and the interface is transparent \cite{BdG-redundancy}.  In each plot, the BdG spectrum
for two values of the superconducting gap are superposed, $\Delta/E_{L}\!=\!0.1$
(green) and $0.5$ (blue). The smaller value is more realistic, but
results for the larger value are instructive because the bulk LLL
QH states are within the superconducting gap, $\Delta>\mu_{\textrm{gr}}=E_{L}/4$,
making the figures clearer. 

For the larger superconducting gap (blue), 
the bulk LLL states at the energies of the electron and hole Dirac points are clear (horizontal blue lines at $\pm \mu_\textrm{gr}$).  The range
of the superconductor's continuum states---the thicket of states at $E>\Delta$---differs for the three cases due to the size
of the Fermi surface. It is largest for dense stitching [panel (a)] in which the
superconductor's band structure is folded back into the graphene Brillouin zone.
For the smaller superconducting gap (green), the QH states hybridize with the superconducting
continuum and largely disappear as a distinct signature in the spectrum.
 
Two CAEM are present at the chemical potential near $k\!=\!K$ and $K'$,
while the two terminated-lattice states on the other edge of
the nanoribbon are unchanged [compare to Fig.\,\ref{fig:spectrum_term-latt}(c)].
The velocity of the CAEM are substantially reduced compared to a terminated-lattice edge, especially for the smaller 
$\Delta$ (see insets). This velocity reduction results from the slow
decay of the wavefunction in the superconductor which implies that
the particles spend a substantial amount of time there. The velocity
reduction is consistent with the QH rule of thumb \cite{[{}][{ p.\,311.}]{IhnBook}} that a softer 
barrier leads to a smaller velocity. 

A striking feature of the CAEM dispersion in all three cases is that
the $E=0$ state is very close to the center of the valley, $k_0\!\approx\!K$ or $K'$. 
Indeed, the dispersion is \emph{valley degenerate} \citep{AkhmerovValleyPolarPRL07}, by which we mean 
\begin{equation}
E_{K}(\delta k)=E_{K'}(\delta k)\label{eq:valley-degen}
\end{equation}
where $\delta k$ is the deviation from the corresponding valley center.
Combining this with BdG particle-hole symmetry 
$E_{K}(\delta k)=-E_{K'}(-\delta k)$, one finds that the dispersion is antisymmetric within each valley, 
\begin{equation}
E_{K}(\delta k)=-E_{K}(-\delta k) \;. \label{eq:valley-antisymm}
\end{equation}
The excellent antisymmetry shown in the insets thus demonstrates nearly perfect valley degeneracy.

Furthermore, we find that a valley degenerate spectrum \emph{requires}
a transparent, well-matched interface. The breaking of valley degeneracy is readily
seen by placing a barrier at the interface (the effect of
changing $\mu_{S}$ is discussed in the next section). In the tight-binding
model, the hopping at the interface, $t_{NS}$, is reduced, and we
presented above in Fig.\,\ref{fig:TBdispersion-tNS-1} results for
the larger value of $\Delta$. In the continuum model, the parameter
controlling the transparency of the interface is the strength of the
$\delta$-function potential at the interface, $V_{0}$, which affects
the boundary condition Eq.\,(\ref{eq:V0-bndcond}) by effectively
increasing the imaginary part of $q$ (see App.\,\ref{sec:CAES-Calc}
for details). 

\begin{figure}
\includegraphics[width=2.7in]{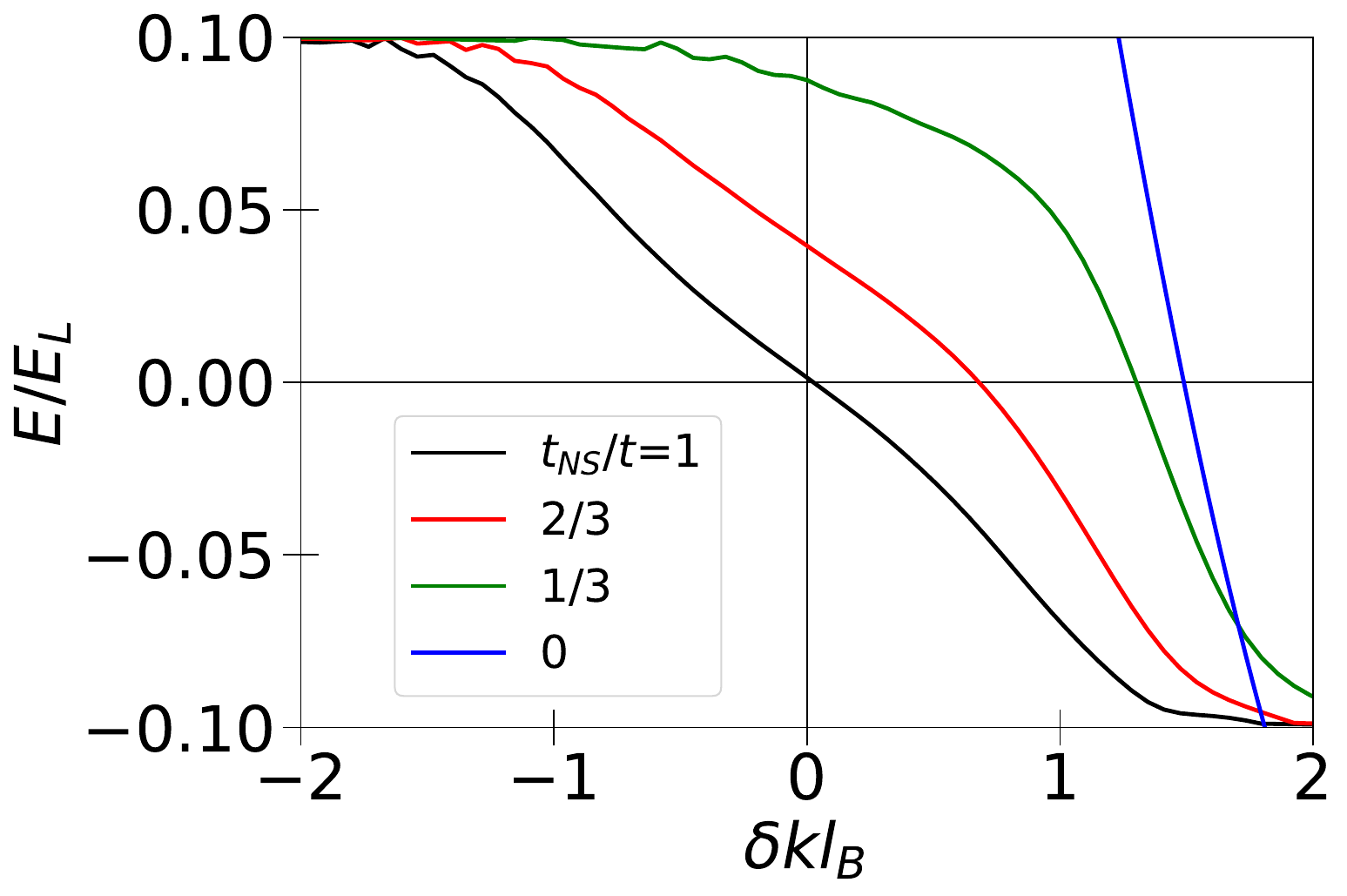}
\caption{Edge mode dispersion for different strengths
of coupling to the superconductor: tight-binding results for small
pairing gap, $\Delta/E_{L}\!=\!0.1$ ($\Delta\!<\!\mu_\textrm{gr}$), and zoom on
the $K$ valley. Here, $E\!<\!\Delta$ and  $\delta k\!\equiv\! k-K$ is scaled by $l_{B}$ [compare to the
continuum result in Fig.\,\ref{fig:MainResults}(c)]. The values
$t_{NS}/t\in\{0,1/3,2/3,1\}$ correspond to interface transparency
$\{0,0.25,0.7,0.85\}$, respectively \cite{Transparency-def}. The spectrum evolves from the
terminated-lattice edge mode (blue) to the fully-hybridized interface mode
centered on the $K$ point (black). (Sparse stitching, otherwise standard parameters, Sec.\,\ref{subsec:Parameters}.)}
\label{fig:TBdispersion-tNS} 
\end{figure}

Figs.\,\ref{fig:TBdispersion-tNS} and \ref{fig:MainResults}(c)
show tight-binding and continuum results, respectively, for the CAEM
dispersion in the $K$ valley for the smaller value of the gap, $\Delta=0.1E_{L}$.
The avoided crossing for small $t_{NS}$ is less clear than in Fig.\,\ref{fig:TBdispersion-tNS-1} because of hybridization with
the continuum of states in the superconductor for $E>\Delta$. 
Nevertheless, in both the
tight-binding and continuum results, there is a similar evolution
from the QH edge state to an antisymmetric dispersion about the center
of the valley. We note an interesting change in the curvature of the
dispersion in both the tight-binding and continuum results: for a
low transparency interface, the curvature is negative near the $K$
point, while for full-transparency, there is an inflection point at
$\delta k=0$ and the curvature is positive for $\delta k<0$. 

In these zooms on the $K$ valley, the decrease in velocity of the
CAEM as transparency increases is particularly clear. We discuss
this further in App.\,\ref{subsec:Vg_appendix}
for the continuum model, giving an analytic expression.
Over an interval $\delta k$ of order $1/l_{B}$, the energy changes
by $\sim\!E_{L}$ for a QH edge state but only by $\sim\!\Delta$
for a CAEM. 

\begin{figure}
\includegraphics[width=2.5in]{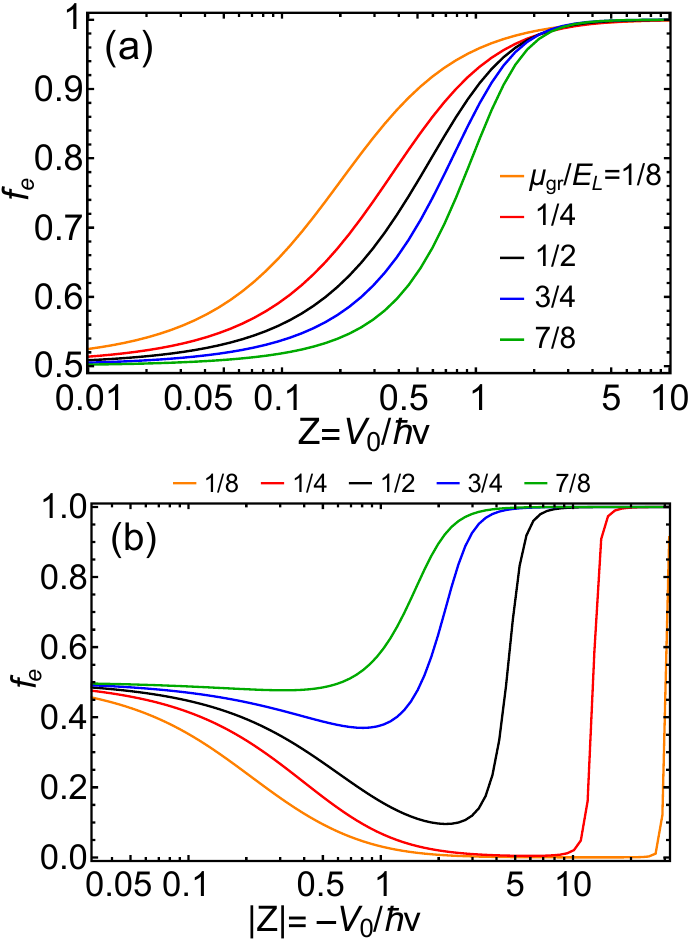}
\caption{Fractional electron content in graphene [$e$-$h$ hybridization, Eq.\,(\ref{eq:fe-define})]
as a function of the strength of (a) a barrier or (b) a potential
well at the interface. Continuum model results are shown, 
varying the filling, $\mu_{\textrm{gr}}/E_{L}$, of the CAEM. 
(Standard parameters, Sec.\,\ref{subsec:Parameters}.)}
\label{fig:ElecFrac-vs-Z}
\end{figure}

Hand in hand with these changes in the CAEM dispersion, the presence
of a barrier changes the electron-hole hybridization of the state
at the chemical potential. Fig.\,\ref{fig:ElecFrac-vs-Z} shows the
hybridization as a function of the barrier strength in the continuum
model ($Z\!=\!V_0 /\hbar v$), for the case of perfect hybridization ($\widetilde{v}^{2}\!=\!1$) when $Z\!=\!0$. The presence
of a barrier at the interface prevents contact with the superconductor
and thus drives the $K$-valley mode toward being fully electron-like
[panel (a)]. The effect of the barrier is more pronounced for
lower filling where the particles have less kinetic energy. On the
other hand, panel (b) shows that a potential well at the interface,
$Z<0$ with $|Z|\sim1$, can enhance the hole content.

\subsection{Valley Degeneracy, Interface Transparency, \\and Andreev Reflection\label{subsec:ValleyDegen-Transp-AR}}

Valley degeneracy is \emph{not} accidental but rather is a universal
feature of well-hybridized states \citep{AkhmerovValleyPolarPRL07}. Moreover,
valley degeneracy is \emph{independent} of the graphene chemical potential
(for $\mu_{\textrm{gr}}$ not too close to the Dirac point): the interface
mode ``self-aligns'' so that $E_{K}(\delta k\!=\!0)\!=\!0$ throughout the
LLL [shown explicitly for the continuum model in Fig.\,\ref{fig:MainResults}(b)]. We find that this holds for \emph{all} of the well-matched, highly
transparent tight-binding and continuum cases that we have studied. 

There is a natural connection between strong Andreev reflection, good
interface transparency, and valley degeneracy. Since $e$-$h$ conversion
can occur only in the superconductor, an incoming electron should
interact with the superconductor as much as possible. A transparent
interface ensures that any ingoing electron component from the graphene
comes out as a hole (not as a scattered electron). Furthermore, since
the electron and hole components in the superconductor have equal
weight at $E=0$, so should the two components in graphene
in order to achieve good wavefunction matching at the interface. In
fact, the two components should have the same spatial profile (as a function of $y$): a Fourier decomposition of the incoming
electron leads to a hole emerging with the same Fourier decomposition
(up to phases) and so the same shape. Thus, for strong Andreev reflection,
the guiding center coordinate of the electron and hole components
of the $E\!=\!0$ state must be similar. 

On the other hand, the wavevector along the interface of the electron
and hole components must be equal. 
But for electrons and holes the relation between the guiding center and the wavevector is
opposite: $\textrm{sign}(\delta \overline{y})\!=\!-\textrm{sign}(\delta k)$
for electrons but $\textrm{sign}(\delta \overline{y})\!=\!+\textrm{sign}(\delta k)$
for holes (with our sign conventions). To have the same guiding
center \emph{and} wavevector requires, then, $\delta \overline{y}\! =\!\delta k\!=\!0$. 

In the LLL of the terminated lattice spectrum, Fig.\,\ref{fig:spectrum_term-latt},
no such a state exists.  In contrast, for a transparent interface, $t_{NS}\sim t\gg E_{L}\gg\Delta$,
an approximately Gaussian wavefunction centered at $\delta \overline{y}_0\!=\!0$
can be constructed, by mixing in small amplitudes of higher Landau level edge states. Alternatively, using the bulk QH complete basis,
a small modification of the bulk LLL wavefunction at $\delta \overline{y}\!=\!\delta k\!=\!0$ triggered by $t_{NS}$
shifts its energy from the Dirac point to the chemical potential.  In this way, a large interface matrix element $t_{NS}$ can produce \textit{both} an electron and a hole at the chemical potential centered on the interface, which are then coupled by the pairing matrix element $\Delta$. 
\emph{Thus, we see that for a transparent interface a ``self-aligned''
$e$-$h$ interface mode, i.e.\ $\delta \overline{y}_{0}\!=\!\delta k_0 \!=\!0$ with no tuning, is natural.}

This self-aligned interface mode is evident in the tight-binding numerics,
Figs.\,\ref{fig:psi-tightbinding}, \ref{fig:Spectra-matched}, and \ref{fig:TBdispersion-tNS} and Table \ref{tab:T-fe-dk_forFigs},
as well as the continuum results, Fig.\,\ref{fig:MainResults}. Once $E(\delta k\!=\!0)=0$ is established, full valley degeneracy
follows in two steps. First, the spectrum within a given valley should
be antisymmetric in $\delta k$ because the role of electrons and
holes is exactly conjugate for a transparent interface. Second,
the BdG particle-hole symmetry then implies valley degeneracy. 

The general argument given above manifests itself
clearly in the equations for the continuum solution in App.\,\ref{sec:CAES-Calc}.
Perfect hybridization means $|C_{1}|\!=\!|C_{2}|$ [see Eq.\ (\ref{eq:psiGrK})].
Since the ratio of these two coefficients, given in
Eq.\,(\ref{eq:AppA_C1dC2}), depends on the ratio of closely related
parabolic cylinder functions evaluated at $y\!=\!0$, one can readily
deduce that $\left|C_{1}/C_{2}\right|\!=\!1$ at $E\!=\!0$ implies $\delta \overline{y}_{0}\!=\!\delta k_0 \!=\!0$ independent of $\mu_\textrm{gr}$: this is the self-aligned CAEM. 
The secular equation then gives full valley degeneracy.

\section{Vary Fermi Energy of Superconductor\label{sec:Vary-E_Fsup}}

In this section, we return to an issue mentioned briefly above:
how the electron-hole hybridization is influenced by the superconductor's Fermi energy. 

\subsubsection{Tight-binding model \label{subsubsec:Xmatch-tbind}}

\begin{figure}
\includegraphics[width=3in]{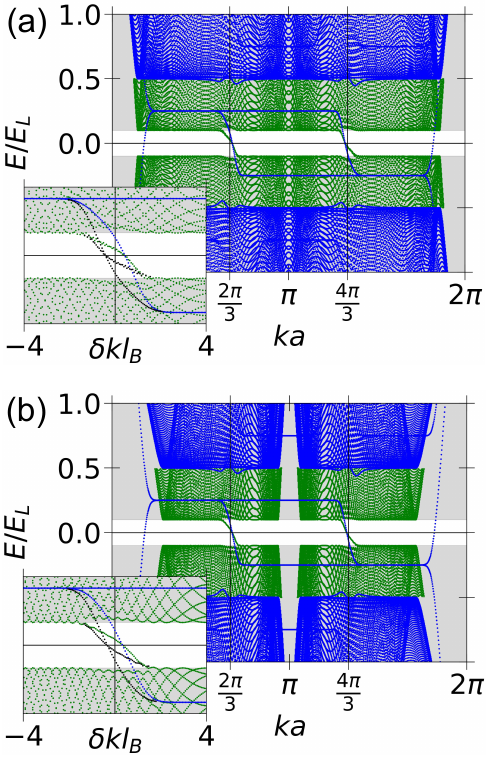}
\caption{Energy of BdG states for larger Fermi surface in the superconductor: (a)\ sparse square lattice with $\mu_{S}/t=4.2$ and (b)\ honeycomb lattice with $\mu_{S}/t=0.8$.  Spectra for two values of the superconducting gap are superposed, $\Delta/E_{L}\!=\!1/10$ (green) and $1/2$ (blue). 
Insets: Zoom on the region around the $K$ point in which $-E_{K}(-\delta k)$ is also plotted (black). The black data do not overlay the blue and green---the deviation of $k_0$ from the valley center is given in Table~\ref{tab:T-fe-dk_forFigs}---demonstrating that valley degeneracy is broken. (Parameters otherwise standard, Sec.\,\ref{subsec:Parameters}.)}
\label{fig:E_largeFsurf}
\end{figure}

\begin{figure}
\includegraphics[width=2.5in]{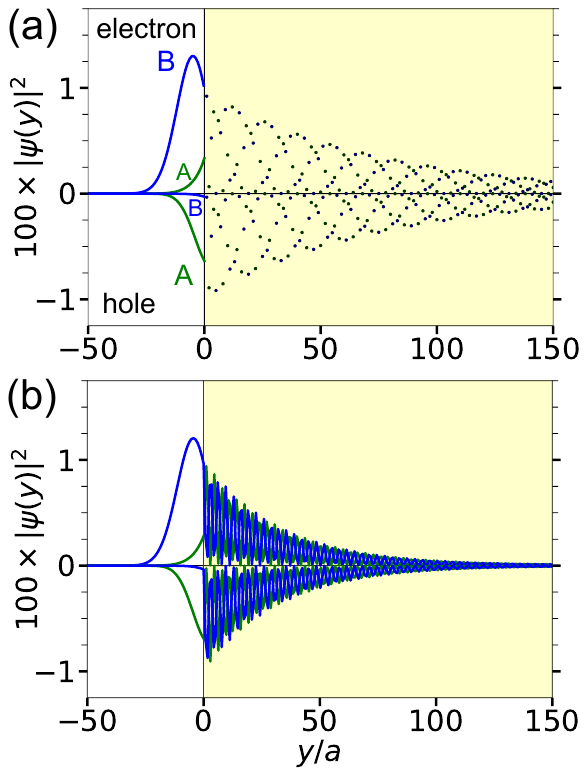}
\caption{Mode wavefunctions $|\psi(y)|^{2}$ of the $E\!=\!0$ state in the $K$ valley for the spectra in Fig.\,\ref{fig:E_largeFsurf}: (a)\  sparse square lattice with $\mu_{S}\!=\!4.2\,t$ and (b)\ honeycomb lattice with $\mu_{S}\!=\!0.8\,t$ ($\Delta/E_{L}\!=\!1/10$). The degree of $e$-$h$ hybridization is given in Table \ref{tab:T-fe-dk_forFigs}.  The wavefunction has the same qualitative structure as in the well-matched cases (Fig.\,\ref{fig:psi-tightbinding}), but the hole content is substantially smaller than the electron content. (Parameters otherwise standard, Sec.\,\ref{subsec:Parameters}.)}
\label{fig:psi_largeFsurf}
\end{figure}

In the tight-binding model, suppose we make the size of the Fermi surface larger and so more metallic-like. For the honeycomb lattice, we increase $\mu_{S}$ substantially, and for the sparse square lattice, we bring $\mu_{S}$ closer to the middle of the band. Since the dense square lattice has a Fermi surface that covers the entire Brillouin zone (due to  folding), we do not consider it for this comparison. The resulting spectra are shown in Fig.\,\ref{fig:E_largeFsurf} and wavefunctions in Fig.\,\ref{fig:psi_largeFsurf}. By comparing the widths of the superconductor continuum to those in Fig.\,\ref{fig:Spectra-matched}, one can see the extent to which the Fermi surface has been enlarged. 

The spectra here are qualitatively similar to each other but distinctly different from the optimal case (Fig.\,\ref{fig:Spectra-matched}) in that valley degeneracy is absent. Valley degeneracy implies antisymmetry, Eqs.\,(\ref{eq:valley-degen})-(\ref{eq:valley-antisymm}), and the black lines in the insets [which are $-E_{K}(-\delta k)$] clearly do not overlay the blue and green.  
Quantitatively, the deviation $\delta k_0$ of the $E\!=\!0$ state from $K$ is given in Table \ref{tab:T-fe-dk_forFigs} for both cases: though much larger than for the well-match\-ed cases, these values are still smaller than for the QH terminated-lattice edge state 
\footnote{In the continuum model, such a deviation of the propagating mode at
the chemical potential, $\delta k_0$ is obtained for $\tilde{v}^{2}=0.38$ and $0.41$, respectively.}. 
Because $\delta k_0$ is of order $1/l_{B}$, the guiding center coordinates of the electron and hole components differ by of order $l_B$, and thus their spatial profiles differ substantially.

The differing spatial profiles are clear in the wavefunctions in Fig.\,\ref{fig:psi_largeFsurf}. On the QH side, the electron weight is considerably larger than the hole weight---quantitatively, see Table \ref{tab:T-fe-dk_forFigs} for the values of $f_{e}$. The form of the electron component on the B sublattice continues to be truncated bulk-like.  
On the superconductor side, the wavefunction is no longer as simple as in Fig.\,\ref{fig:psi-tightbinding}: the rapid oscillations due to the large $k_{FS}$ are out of phase with the lattice.

\begin{figure}
\includegraphics[width=3.0in]{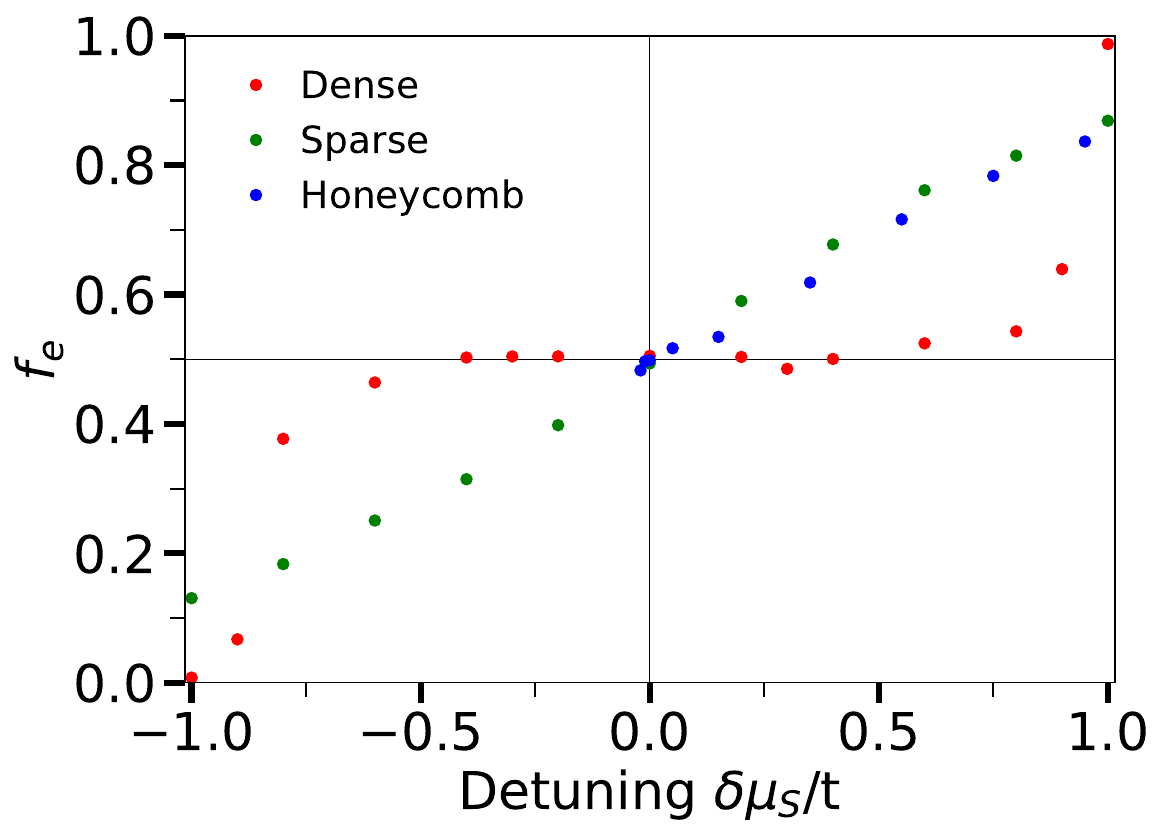}
\caption{Fractional electron content in graphene [$e$-$h$ hybridization, Eq.\,(\ref{eq:fe-define})] as a function of the chemical potential of the superconductor
for the three tight-binding stitchings (Fig.\,\ref{fig:lattices}). Detuning is from the optimal value (see
Sec.\,\ref{subsec:Parameters}; parameters otherwise standard). Note
the especially wide window of excellent hybridization in the dense case (red).}
\label{fig:fe-vs-muS_tbind}
\end{figure}

Importantly, in order to achieve good electron-hole hybridization, $\mu_{S}$ does \emph{not} have to be very carefully chosen. Fig.\,\ref{fig:fe-vs-muS_tbind} shows how the hybridization changes upon detuning $\mu_{S}$ from the optimal value. As a rough criterion for good hybridization, we use $|f_e -\frac{1}{2}| \!\alt\! 0.1$. The wavefunction at the chemical potential 
satisfies this criterion for $\mu_{S}\!\in\![3.3,4.8]t$ for dense stitching, $\mu_{S}\!\in\![4.8,5.2]t$ for sparse stitching, and $\mu_{S}\!\in\![0,0.4]t$ for the honeycomb lattice. 
For both square lattice cases, the regime of good matching (strong coupling) occurs when the Fermi energy of the superconductor is much larger than that of graphene, suggesting that graphene is particularly suitable for interfaces to superconductors. 

Dense stitching, which is especially attractive since the Fermi surface is large, 
yields good $e$-$h$ hybridization for about one fifth of the total band width of the superconductor---a very substantial range. Because in the experimental systems the superconductor likely connects to both sublattices of graphene and has several Fermi surface sheets, we expect the dense stitching results to be the most relevant. 

We find that the degree of hybridization depends only weakly on graphene doping, 
$\mu_{\textrm{gr}}/E_{L}$, in all of these cases (not shown). 
Thus, if the superconductor is well matched to graphene (i.e.\ with $\mu_{S}$ in the optimal range) then, 
as in the continuum model Fig.\,\ref{fig:MainResults}(b), there is good electron-hole hybridization throughout the LLL.

\subsubsection{Continuum model \label{subsubsec:Xmatch-cont}}

\begin{figure}[b]
\includegraphics[width=2.6in]{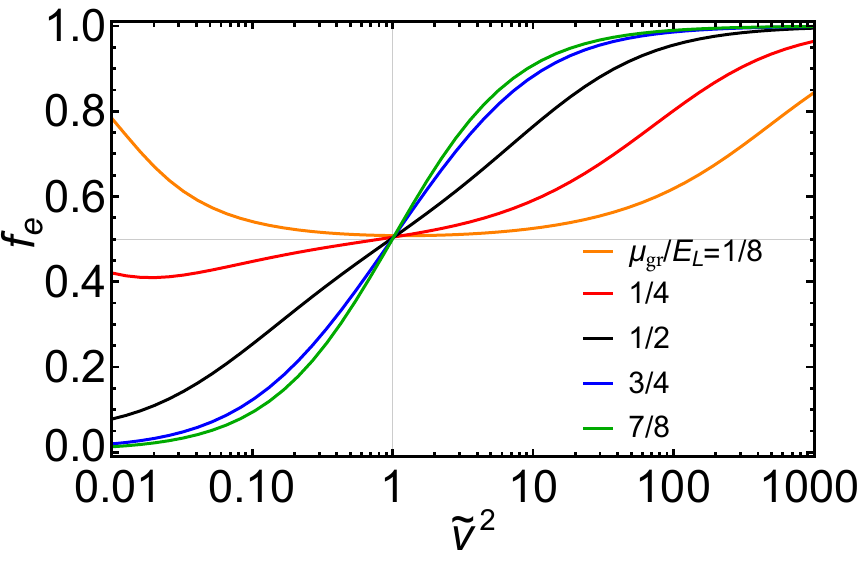}
\caption{Fractional electron content in graphene 
as a function of imperfect matching due to the superconductor
chemical potential (no barrier, $Z\!=\!0$). Continuum
model results shown for different CAEM filling $\mu_{\textrm{gr}}/E_{L}$. 
$\widetilde{v}^{2}$ is defined in Eqs.\,(\ref{eq:def-vtilde})-(\ref{eq:vtildesq1}).
(Parameters otherwise standard, Sec.\,\ref{subsec:Parameters}.)}
\label{fig:ElecFrac-vs-mu}
\end{figure}

How the hybridization changes with $\mu_{S}$ for the continuum
model is shown in Fig.\,\ref{fig:ElecFrac-vs-mu}.
As the Fermi energy of the superconductor deviates from its optimal
value, $\widetilde{v}^{2}\lessgtr1$, the CAEM has more
electron or hole character. For low graphene
filling, the insensitivity of the hy\-bridization to the superconductor's properties
stands out. 

Even for a large variation in the properties of the superconductor, Fig.\,\ref{fig:MainResults}(b)
shows that the hybridization remains good throughout the LLL.
The underlying reason is that current continuity involves
the velocity at the Dirac point, which is insensitive to external
parameters. For instance, for a given superconductor, the velocity
$\hbar K/m$ is fixed with respect to the Dirac point velocity.
For $\hbar K/m=2v$, varying $\widetilde{v}^{2}$ from $1/2$ to $2$ 
[as in Fig.\,\ref{fig:MainResults}(b)]
corresponds to changing the Fermi energy of the superconductor by
a factor of $2$, while for $\hbar K/m\ll v$, such a variation changes
$\mu_S$ by a factor of $4$. Only the unlikely case $\hbar K/m\approx v$
leads to undesirable rapid variability. 
\emph{Thus, in contrast to the usual semiconductor case 
\citep{HoppeZulickePRL00, GiazottoPRB05, KurilovichDisorderAES_NCom23, MichelsenSchmidt_SupercurEnable_PRR23, ADavid_Grenoble23},
no fine-tuning is needed to have excellent Andreev reflection in graphene in the QH regime.}

\begin{figure}
\includegraphics[width=2.5in]{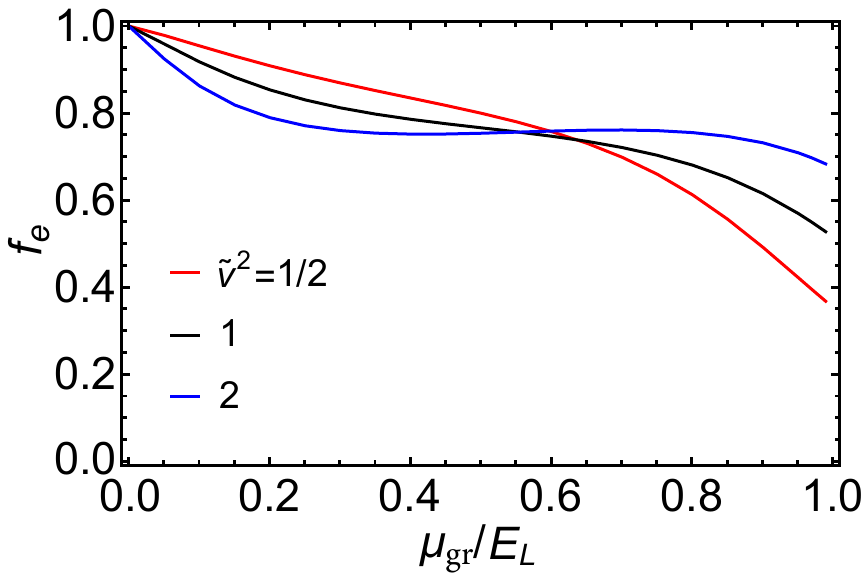}
\caption{Fractional electron content in graphene
as a function of the filling of the LLL interface mode for an imperfect
interface. Shown for three values of the superconductor Fermi energy,
$\widetilde{v}^{2}\!=\!1/2,1,2$, in the continuum model with the interface
barrier strength set to $Z\!=\!V_{0}/\hbar v\!=\!0.5$. (Parameters otherwise standard, Sec.\,\ref{subsec:Parameters}.)}
\label{fig:ElecFrac-Z0.5}
\end{figure}

Finally, we present in Fig.\,\ref{fig:ElecFrac-Z0.5} results when both a detuned $\mu_{S}$ 
and a small interface barrier, $Z=0.5$, are present.
There continues to be only a modest dependence on the Fermi energy of
the superconductor through $\widetilde{v}^{2}$, but there is somewhat more
dependence on the graphene chemical potential $\mu_{\textrm{gr}}$.
Note, however, that the dependence is smooth:
there continues to be no indication that fine-tuning is needed to obtain 
strong Andreev reflection. 

\section{Spin: Zeeman-Split Modes \label{sec:Spin-Zeeman}}

We now present results in which the Zeeman effect is included (spin-orbit coupling is still not included). 
The magnetic field in graphene acts on the spin 
of the electron and splits the spin up and down solutions, 
leading to two modes in each valley \cite{fn:SpinConvention}. 
The modes in the $K$ valley are, of course, related to those in the $K'$ valley by the BdG particle-hole 
symmetry: $E_{K\uparrow}(\delta k)=-E_{K'\downarrow}(-\delta k)$. 

\begin{figure}
\includegraphics[width=3in]{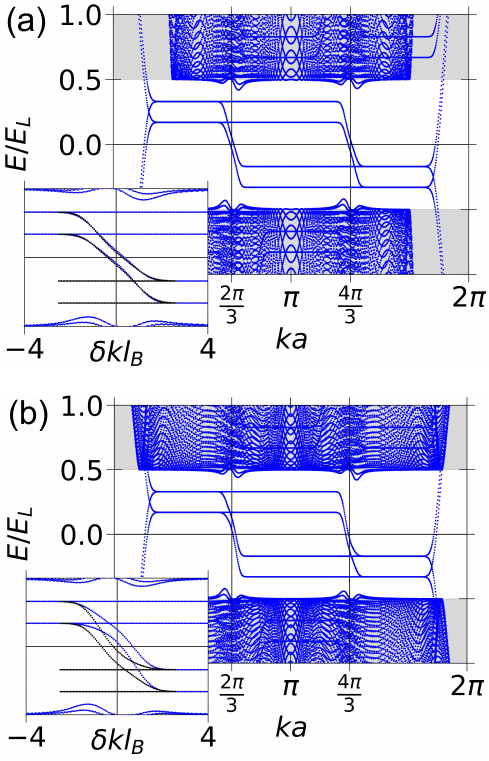}
\caption{Energy of BdG states for sparse stitching with Zeeman splitting: (a) well-matched, transparent interface ($\mu_{S}/t\!=\!5$) and (b) imperfect ($\mu_{S}/t\!=\!4.2$). 
Insets: Zoom on the region around the $K$ point in which the $K'$-valley spectrum, $E_{K'}(\delta k)$, is also plotted (black). The deviation of $k_0$ from the valley center is given in Table \ref{tab:T-fe-dk_Zeeman}. Valley-degeneracy is very good for the transparent interface (though the dispersion is not antisymmetric about $\delta k\!=\!0$) but broken for a non-ideal interface.  ($\Delta/E_{L}\!=\!0.5$; parameters otherwise standard, Sec.\,\ref{subsec:Parameters}.)}
\label{fig:Spin-Spectrum}
\end{figure}

The spectrum for tight-binding with sparse stitching is shown in Fig.\,\ref{fig:Spin-Spectrum}(a) for a well-matched case. The Zeeman splitting is $\Delta_z=E_L/6$ (corresponding to a large g-factor for illustration).
The bulk LLL is split in two (horizontal lines), and a CAEM connects the higher energy electron line to the higher energy hole line [compare to Fig.\,\ref{fig:Spectra-matched}(b) for zero Zeeman effect].

\begin{table}[b]
\begin{tabular}{c c l}
\parbox[t]{0.5in}{Fig. (panel)} & \parbox[t]{0.72in}{Fractional electron content, $f_{e}$} & 
\parbox[t]{0.98in}{Deviation from valley center, $\delta k_0\,l_{B}$} \\
\hline 
17(a)  & 0.34 & $-$0.28 [unsplit: 0.01]\tabularnewline 
17(b)  & 0.68 & \;\;\,0.36\tabularnewline 
18(a)  & 0.72 & \;\;\,0.17 [unsplit: 0.42]\tabularnewline
18(b)  & 0.89 & \;\;\,0.70\tabularnewline
\hline 
\end{tabular}\smallskip
\caption{\label{tab:T-fe-dk_Zeeman}
For the Zeeman-split tight-binding wave functions in Figs.\,\ref{fig:Spin-Psi-1} and \ref{fig:Spin-Psi-2}, we give the fractional electron content in graphene and the deviation of 
the wavevector $k_0$ for the $E=0$ mode from the center of the $K$ valley: $\delta k_0\,l_{B}\equiv(k_0-K)l_{B}$.}
\end{table}

Importantly, note that the 
spectrum remains \emph{valley degenerate}: the inset highlights $E_{K\uparrow}(\delta k)\!=\!E_{K'\uparrow}(\delta k)$. 
However, because of the spin index, particle-hole symmetry plus valley degeneracy no longer imply that the 
dispersion is antisymmetric within a valley, $E_{K\uparrow}(\delta k) \!\neq\! -E_{K\uparrow}(-\delta k)$. 
Thus, the zero energy excitations need not occur at the valley centers. In fact, the two 
solutions are shifted oppositely from the valley center---the values of $\delta k_0$ are given in Table 
\ref{tab:T-fe-dk_Zeeman}, together with those for no Zeeman effect (labeled unsplit). Though additional momentum 
differences among the $E\!=\!0$ modes are thus generated by Zeeman splitting, no new interference effects 
can occur in transport since the up and down spin modes are completely uncoupled.  

For an imperfectly matched interface, the spectrum is shown in Fig.\,\ref{fig:Spin-Spectrum}(b).
The spectrum now is definitely \emph{not} valley degenerate, just as in the absence of Zeeman splitting
(Fig.\,\ref{fig:E_largeFsurf}): the difference
in the $K$ and $K'$ valley spectra is seen clearly in the inset. As in the transparent
interface (see Table \ref{tab:T-fe-dk_forFigs}), the two modes shift oppositely from the zero-Zeeman case,
but now both modes occur on the same side of the valley center---both $\delta k_{0}$ are positive
in the $K$ valley and negative in $K'$ (see Table \ref{tab:T-fe-dk_Zeeman}).
Since the modes that can interfere now have different $\delta k_{0}$, new interference effects are possible. 

\begin{figure}
\includegraphics[width=2.5in]{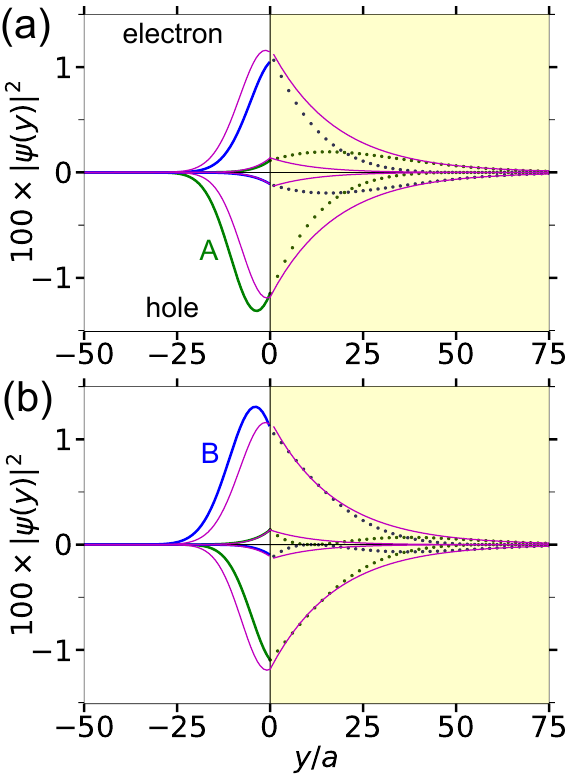}
\caption{Wavefunctions, $|\psi(y)|^{2}$, for the two
Zeeman-split modes in the $K$ valley at $E=0$:
(a) spin up and (b) spin down. The sparse-stitching interface is well matched and so maximally
transparent. The degree of $e$-$h$ hybridization is given in Table \ref{tab:T-fe-dk_Zeeman}. The magenta line is the zero-Zeeman case for
comparison. [Correspond to spectrum in Fig.\,\ref{fig:Spin-Spectrum}(a).]}
\label{fig:Spin-Psi-1}
\end{figure}

Our results for a well-matched, transparent interface are quite different from those
in a recent paper, Ref.\,\citep{CuozzoRossiPRB24}. There, the authors
emphasize that in breaking the SU(4) symmetry of the problem, a Zeeman
field produces modes with different group velocities. We see here
that this is not the case for a transparent interface. Rather, for
a transparent interface, the spectrum remains valley degenerate despite the Zeeman field, and the velocities of all four modes
are the same. On the other hand, our results for an imperfect, poorly matched interface are in agreement with those of Ref.\,\citep{CuozzoRossiPRB24}. 

Ref.\,\citep{CuozzoRossiPRB24} makes an interesting comment on the
possibility of achieving Majorana modes when the canted antiferromagnet
state of the LLL \citep{NomuraMacDonald_QHferro-Gr_PRL06,Goerbig_Gr-StrongB_RMP11,Young_QHferromag-Gr_NatPhy12} is coupled to a superconductor. The authors point
out that the induced gap that supports Majorana modes vanishes if
$\delta k_0$ is too large, and therefore suggest that the observability
of such modes may be overstated in the literature. However, this conclusion
was reached by considering an interface with poor transparency. Here,
we point out that for a well-matched, transparent interface, one finds $\delta k_0  l_B \!\ll\! 1$
in the absence of Zeeman splitting, and even with Zeeman
splitting, $\delta k_0  l_B$ is not large. Thus, the more optimistic scenario for
achieving Majorana modes appears to be possible as long as the graphene-superconductor
interface is transparent. 

\begin{figure}
\includegraphics[width=2.5in]{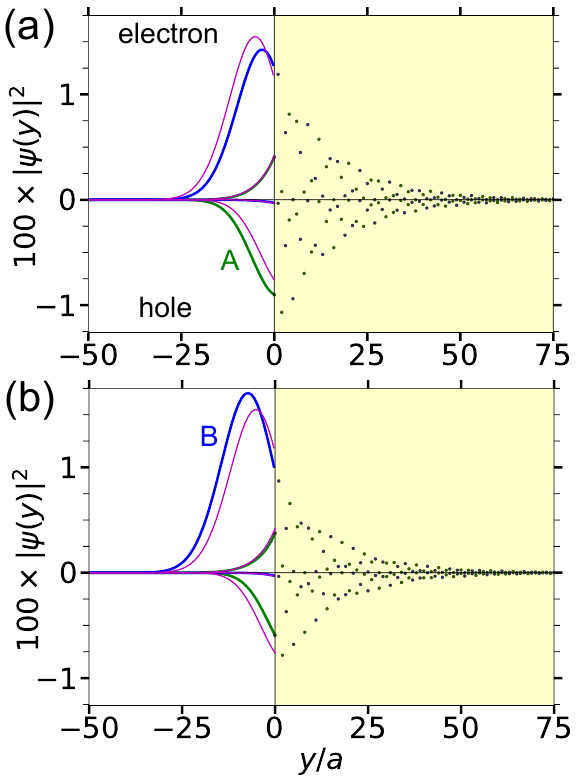}
\caption{For an imperfect interface, wavefunctions
for the two $E\!=\!0$ modes in the $K$ valley as
they are split by the Zeeman effect: (a)~spin up and (b) spin down.
Here, $\mu_{S}/t=4.2$.
The magenta line
is the zero-Zeeman case for comparison (not plotted in superconductor
for clarity). [Correspond to spectrum in Fig.\,\ref{fig:Spin-Spectrum}(b).]}
\label{fig:Spin-Psi-2}
\end{figure}

Turning to the corresponding wavefunctions for $E\!=\!0$, we see in Fig.\,\ref{fig:Spin-Psi-1} that for a transparent interface the elec\-tron-hole balance has been disrupted. Specifically, on the graphene side the main parts of the electron and the hole components, which correspond to the bulk states, displace oppositely relative to the interface, as expected for $\delta k_0\neq0$ (see Sec.\,\ref{subsec:ValleyDegen-Transp-AR}).  
In one mode the electron component shifts toward the interface and gets smaller and the hole component shifts away and gets bigger, while the reverse occurs in the other mode.  In contrast, the small, exponentially decaying surface components are not modified. In the superconductor, the wavefunction continues to have a clear even-odd and electron-hole alternation, though the form is modified. 

For an imperfectly matched interface, the wavefunctions are shown in Fig.\,\ref{fig:Spin-Psi-2}. The effect of the Zeeman field is 
similar to the transparent case (again the exponentially decaying surface components are barely affected). Note that for the mode in panel (a), because the imperfect interface by itself produces $e$-$h$ unbalanced modes, the Zeeman effect actually brings the electron and hole closer spatially and makes their weights more similar. [The opposite is true for the other mode, panel (b).] Thus, under the symmetry breaking connected with the Zeeman effect, Andreev reflection can become \textit{stronger}.

\section{Conclusions \label{sec:Conclusions}}

We have presented results for an infinite zigzag gra\-phene/superconductor
interface \cite{ArmResults} using continuum and tight-binding models in the standard
mean-field approach to spatially inhomogeneous superconductivity (BdG equation). 
Our main assumptions are: (i) disorder is absent and (ii) the fields $\Delta$, $B$,
and $U$ are taken to be step functions at the interface. The continuum results are 
particularly helpful in exploring a wide range of parameters. 

The primary conclusion is that strong Andreev reflection, and hence
strong electron-hole hybridization, is much easier to achieve with
graphene than with a typical semiconductor 2D electron
gas. The reason is that the velocity at the Dirac points in graphene
is large (of order the Fermi velocity of metals), even though the conduction electron density (and hence $k_F$) is low \cite{v-parabolic}.  Furthermore, the Dirac point velocity is independent
of the graphene chemical potential or density. Thus, current continuity
across the interface can be satisfied without a great deal of electron reflection. 

In order to actually realize strong Andreev reflection, the interface
should be fairly transparent. In particular, there should be no obvious barrier, and the superconductor
should be well-matched, having Fermi level states with wavevector
larger than $K$ and velocity of order $v$ (the wave vector and velocity of the Dirac point). This does not appear
to be a stringent restriction, since the Fermi surface
of metals used for thin-film superconductors often have several sheets, allowing for several possible matches. 

Previous work has emphasized that fine-tuning is necessary in order
to achieve electron-hole hybridization at a clean QH/superconductor
interface \citep{GiazottoPRB05,KurilovichDisorderAES_NCom23,MichelsenSchmidt_SupercurEnable_PRR23};
for instance, the chemical potential in the QH material should be
tuned on an energy scale less than $E_L$. We emphasize
that this is \emph{not} the case for graphene. Once one obtains a
fairly well-matched transparent interface, which may involve ``coarse-tuning''
the choice of superconductor on an energy scale of the bandwidth ($\gg\! E_L$),
the $e$-$h$ hybridization in a graphene CAEM is ideal and independent
of filling. 
Indeed, the dispersion of the CAEM self-aligns to be antisymmetric
within each valley, independent of filling. Thus, the spectrum is valley degenerate, and the guiding center
coordinate of the state at the chemical potential is at the graphene/superconductor
interface.

The CAEM wavefunction in graphene exhibits particularly simple structure
in the case of a well-matched interface. The hole component is nearly
the same as the electron but with the sublattices interchanged.
This pairing of two electrons on \textit{different} sublattices is distinctly
different from that induced by tunnel-coupling to a disordered superconductor
in which two electrons on the \textit{same} sublattice are paired. 

In the spatial profile of the wavefunction, the weight of the electron
component is primarily on the sublattice corresponding to the bulk
LLL state (sublattice-valley locking), and is surprisingly close to
a truncated bulk wavefunction. This demonstrates that the terminated-lattice
QH edge states are a poor starting point for understanding the CAEM
of a transparent interface, an unsurprising result in hindsight in
view of the very different boundary conditions.  Rather, in the case of strong graphene-superconductor coupling ($t_{NS}\!\gg\! E_L\!\gg\!\Delta$), bulk-like electron and hole wavefunctions can be made at $E\!=0$ using $t_{NS}$, which are then coupled by superconductivity ($\Delta$). 

In future work, the role of disorder, both in the superconductor and
along the interface, should be further developed in the context of a transparent gra\-phene/super\-con\-duc\-tor interface (for initial work see Ref.\ \citep{Manesco-QHgrS_SP22} and for the importance
of disorder in the  tunneling regime see Ref.\,\citep{KurilovichDisorderAES_NCom23}). 
We suspect that disorder will decrease the dependence on $\mu_{S}$
in our models and so increase the window of good electron-hole hybridization.
A second clear direction is to include the penetration of the magnetic
field into the superconductor in the form of vortices. It is likely
that tunneling of electrons into states in the vortices is an important
source of loss in the experiments \citep{LingfeiGlebCAESNatPhys20,LingfeiGlebLossPRL23}, 
and theoretical work assessing the importance of this effect has recently
appeared \citep{TangAlicea_VortexEnabled_PRB22,SchillerOreg_FermDissip_PRB23,HuLian-Decohere-X24}. 
Third, the role of spin-orbit coupling should be investigated.  While such coupling is weak in graphene, in several of the superconductors typically used it is strong. 

Finally, perhaps the most straight forward extension of our results is to the properties of nanostructures, which involve scattering between these CAEM at junctions, corners, and quantum point contacts.  This would make contact with transport experiments, typically done in nanostructures \cite{ParkLee_S-QHedge_SciRep17, LeeHarvardNatPhy17, SahuBangalorePRL18, LingfeiGlebCAESNatPhys20, LingfeiGlebLossPRL23}.  A preliminary step toward the experiments can be made using the infinite-interface property $f_e$ calculated here [Eq.\,(\ref{eq:fe-define})].  Suppose the interface is connected to incoming and outgoing graphene terminated-lattice edge states.  If one assumes that the electron and hole amplitudes on the outgoing edge directly reflect the hybridization of the interface states, then the probability that an electron exits in the outgoing edge state is simply $f_e$ and the downstream resistance, $R_d$, follows \citep{LingfeiGlebCAESNatPhys20}:
\begin{align}
P_e=f_e\;,\quad R_d=R_H \frac{2f_e-1}{1-f_e}\;,    
\end{align}
where $R_H$ is the Hall resistance.
However, the assumption is not usually valid: it is known that both valley momentum conservation \cite{AkhmerovValleyPolarPRL07} and scattering at corners \citep{DavidGrenoble_Geometrical_PRB23} are important, even in the clean case.  Having focused on properties of the infinite interface here, we leave consideration of these nanostructure effects to future work.  

\begin{acknowledgments}
We thank Ethan Arnault, Anthony David, Gleb Finkelstein, William Klein, and Lingfei Zhao for helpful discussions. 
The work in the USA was supported in part by the U.S.\ Department of Energy, Office of Science, Office of Basic Energy Sciences, Materials Sciences and Engineering Division, under Award No.\ DE-SC0005237.
A.B. acknowledges support by the DoD National Defense Science and Engineering Graduate Fellowship Program. 
G.Z. acknowledges support from the National Natural Science Foundation of China (Grant No.\ 12374158) and the Innovation Program for Quantum Science and Technology (Grant No.\ 2021ZD0302400). 
\end{acknowledgments}

The data that support the findings of this article are openly available \cite{Data_grS-InfInterface}.

\appendix

\section{Specifics of the Models: \\Tight-Binding and Continuum \label{app:ModelSpecifics}}

In this appendix, we specify the tight-binding and continuum models
presented in Sec.\,\ref{sec:Models}. 

In the \textit{tight-binding model}, the normal state spin-projected hamiltonian of the superconductor is a square lattice in the $y>0$ half-plane,  
\begin{align}
H_{S}(x,y>0)= & -t\underset{\langle ij\rangle}{\sum}\left(|i\rangle\langle j|+\textrm{h.c.}\right)
 -(\mu_{S}-4t)\underset{i}{\sum}|i\rangle\langle i|,\label{eq:Ham-tbS-1}
\end{align}
where $\langle ij\rangle$ denotes nearest-neighbor sites.\ $\mu_{S}$ is the chemical potential (Fermi energy) measured from the bottom of the band. Pairing in the superconductor connects the electron and hole states with the same spin \cite{fn:SpinConvention},
\begin{align}
\Delta(x,y>0)= & \Delta\underset{i}{\sum}\left(|i\rangle_{e}\langle i|_{h}+\textrm{h.c.}\right).\label{eq:Delta-tb-1}
\end{align}
In the absence of spin-orbit coupling and the Zeeman effect,  the problem decouples into
two independent, identical problems, leading to doubly degenerate states. We therefore focus on the spin-projected system. For Zeeman splitting in Sec.\,\ref{sec:Spin-Zeeman}, the two problems are still  independent (but not identical), and the spin index can easily be introduced. 

Graphene is modeled by hopping on a honeycomb lattice with phase factors due to the magnetic field: 
\begin{align}
H_{QH}(x,y\leq0)= & -t\underset{\langle ij\rangle}{\sum}\left(e^{-2\pi i\Phi_{ij}/\Phi_{0}}|i\rangle\langle j|+\textrm{h.c.}\right)
\nonumber \\ & 
-\mu_{\textrm{gr}}\underset{i}{\sum}|i\rangle\langle i|,\label{eq:Ham-tbQH-1}
\end{align}
where $\Phi_{ij}=\int_{\mathbf{r}_{j}}^{\mathbf{r}_{i}}\mathbf{A}\cdot d\mathbf{l}$ and $\Phi_{0}=h/e$ is the flux quantum. We take $\mathbf{A}(x,y)=By\hat{x}\,\theta(-y)$, and with this choice $\Delta$ in Eq.\,(\ref{eq:Delta-tb-1}) is constant and real. $\mu_{\textrm{gr}}$ is the chemical potential (Fermi energy) measured from the Dirac point.  
The terminal row of the zigzag edge is at $y=0$ and part of the A sublattice. The graphene half-sheet has period $a$ in the $x$ direction (next-nearest-neighbor distance).  

To specify the hopping hamiltonian connecting the graphene and superconductor lattices, we use $a$ as the unit of length in writing coordinates for a position space basis $|x,y\rangle$. 
For ``dense'' stitching [Fig.\,\ref{fig:lattices}(a)], we have 
\begin{align}
H_{T,\text{dense}}= & H_{T}^{(\textrm{A})}+H_{T}^{(\textrm{B})}\nonumber \\
\equiv & -t_{NS}\underset{n}{\sum}^{(\textrm{A})}\left[|n,\tfrac{1}{2}\rangle\langle n,0|+\textrm{h.c.}\right]\\
 & -t_{NS}\underset{n}{\sum}^{(\textrm{B})}\left[|n+\tfrac{1}{2},\tfrac{1}{2}\rangle\langle n+\tfrac{1}{2},-\tfrac{\sqrt{3}}{2}|+\textrm{h.c.}\right],\nonumber 
\end{align}
where the first term connects the square lattice and the terminal A sites of graphene and the second term connects the neighboring B sites directly to the square lattice. Second, in the ``sparse'' stitching scenario [Fig.\,\ref{fig:lattices}(b)], the square lattice has the same period as the zigzag edge and only the terminal A sites are connected to the square lattice, 
\begin{equation}
H_{T,\text{sparse}}=H_{T}^{(\textrm{A})}.
\end{equation}
Third, in the ``honeycomb'' scenario, the graphene lattice is simply continued into the superconductor [Fig.\,\ref{fig:lattices}(c)]. 

After constructing the Hamiltonian using the Kwant package \cite{GrothKwantNJP14},
we solve for the band structure by Fourier transforming along the $x$-direction, using the quasi-1D unit cell of a nanoribbon with width defined by $-y_{QH}<y<y_{S}$ 
\cite{WidthFootnote}. The fractional electron content in graphene is given in terms of the wavefunction by
\begin{equation}
    f_e \equiv \frac{\sum_{i \in \textrm{gr}} |\psi_i^e|^2}
    {\sum_{i \in \textrm{gr}} |\psi_i^e|^2 + |\psi_i^h|^2}
    \label{eq:fe-def-tbind}
\end{equation}
where the sum is over all graphene sites 
(both A and B). 

In the \textit{continuum model}, the wavefunction in the superconductor
is a solution to the BdG equation with $H_{S}$ and $\Delta$ in 
Eqs.\,(\ref{eq:HS-contin})-(\ref{eq:Delta_cont}) and 
has the Nambu spinor form 
\begin{align}
\Psi_{S}(x,y\!>\!0) & =e^{ipx}\left\{ C_{3}\Psi_{S,p}^{+}(y)+C_{4}\Psi_{S,p}^{-}(y)\right\} 
\nonumber \\ & 
\equiv e^{ipx}\left(\!\!\begin{array}{c}
\psi_{S,p}^{e}(y)\\
\psi_{S,p}^{h}(y)
\end{array}\!\!\right)\label{eq:psiS}\\
\textrm{with}\;\; & \Psi_{S,p}^{+}(y)=\sqrt{\,q''}e^{iqy}\left(\!\begin{array}{c}
1\\
\gamma
\end{array}\!\right)
\end{align}
and $\Psi_{S,p}^{-}(y) \!=\! [\Psi_{S,p}^{+}(y)]^{*}$ 
(with $\pm$ indicating the  direction of evanescent current perpendicular to the interface) \citep{BTK-PRB82}. 
For $E<\Delta$, $q$ and $\gamma$ are given by
\begin{align}
\gamma & =\frac{E-i\sqrt{\Delta^{2}-E^{2}}}{\Delta},\label{eq:def-Gamma}\\
q^{2} & =(k_{FS}^{2}-p^{2})+i\frac{2m}{\hbar^{2}}\sqrt{\Delta^{2}-E^{2}}. 
\label{eq:def-qsq}
\end{align}
Note that $|\gamma|=1$, implying an equal weight of electrons and holes
in the superconductor. For the $E=0$ states, $\gamma=-i$ and $q^{2}$
simplifies as in Eqs.\,(\ref{eq:q_Eeq0}) and (\ref{eq:vtildesq1}).

On the graphene side, the wavefunction is proportional to $e^{ikx}$,
and because of translational invariance, the amplitudes for the $K$
and $K'$ valleys are independent, Eq.\,(\ref{eq:k-match}). 
In terms of spinors in both sublattice and $e$-$h$ indices, the solution in valley $K$ is of the form 
\begin{align}
&\Psi_{K}(x,y<0) \equiv e^{ikx}\Psi_{K,k}(y), \\
&\Psi_{K,k}(y) \equiv C_{1}\!\left(\!\begin{array}{c}
\phi_{1k}^{e}(y)\\
\phi_{2k}^{e}(y)
\end{array}\!\right)\!\otimes\!\left(\!\begin{array}{c}
1\\
0
\end{array}\!\right)\!  + C_{2}\!\left(\!\begin{array}{c}
\phi_{1k}^{h}(y)\\
\phi_{2k}^{h}(y)
\end{array}\!\right)\!\otimes\!\left(\!\begin{array}{c}
0\\
1
\end{array}\!\right),
\label{eq:psiGrK}
\end{align}
with two undetermined coefficients. 
The transverse wavefunctions appearing in the boundary conditions Eqs.\ (\ref{eq:bndcond1})-(\ref{eq:bndcond2}) are, for example, 
$\psi_{K\textrm{A},k}^e(y)\!=\!C_1\phi_{1k}^{e}(y)$ and $\psi_{K\textrm{B},k}^h=C_2\phi_{2k}^{h}(y)$. 
For the $K'$ valley, the form of the transverse wavefunction is given 
by a simple interchange of the sublattices: $\Psi_{K',k}(y)  = -i\sigma_y \Psi_{K,k}(y)$.

In the absence of a magnetic field, the transverse wavefunctions are
plane waves. An ingoing electron is reflected as either an electron
or a hole (see App.\,\ref{sec:B=0Calc}).

In the QH regime, the transverse solutions $\phi_{n,k}^{\beta}(y)$
are parabolic cylinder functions \citep{HoppeZulickePRL00,AkhmerovValleyPolarPRL07,Romanovsky_DiracQH_PRB11}.
We normalize them to unity, 
\begin{equation}
\int_{-\infty}^{0}dy\{|\phi_{1k}^{\beta}(y)|^{2}+|\phi_{2k}^{\beta}(y)|^{2}\}=1,\label{eq:Norm-phis}
\end{equation}
 and write them explicitly in Appendix \ref{sec:CAES-Calc}.
In addition, we normalize the full transverse wavefunction to unity
\begin{equation}
1= \sum_{i=1}^4 |C_i|^2  = |C_{1}|^{2}+|C_{2}|^{2}+2|C_{4}|^{2}.
\end{equation}

The four unknowns in the wavefunction---the $\{C_{i}\}$ in Eqs.\,(\ref{eq:psiS})
and (\ref{eq:psiGrK})---are determined by the boundary conditions
(\ref{eq:bndcond1})-(\ref{eq:bndcond2}) together with the normalization conditions. Thus, the energies
and wavefunctions of the interface modes are found. In terms of the
wavefunction (\ref{eq:psiGrK}), the fractional
electron content in graphene is 
\begin{align}
f_{e} & \equiv\frac{|C_{1}|^{2}}{|C_{1}|^{2}+|C_{2}|^{2}}.\label{eq:fe-def}
\end{align}

\section{Boundary Condition in the\\Continuum Model \label{sec:BoundaryConds}}

In this appendix we derive the current continuity boundary condition
between graphene and a superconductor with parabolic dispersion, within
the continuum model. We match the expectation value of the particle
current (not the charge current). Since electrons cannot be instantaneously converted
into holes, the current of electrons and holes must each match. 

The expectation value of the particle current is conveniently written in terms of the components of the Nambu spinor. 
In the superconductor, in terms of the components defined in Eq.\,(\ref{eq:psiS}), the current 
perpendicular to the interface is 
\begin{subequations}
\begin{align}
j_{S}^{e} & =\frac{i\hbar}{2m}\left(\psi_{S}^{e}\partial\psi_{S}^{e*}-\psi_{S}^{e*}\partial\psi_{S}^{e}\right),\\
j_{S}^{h} & =-\frac{i\hbar}{2m}\left(\psi_{S}^{h}\partial\psi_{S}^{h*}-\psi_{S}^{h*}\partial\psi_{S}^{h}\right),\label{eq:jhole_sup}
\end{align}
\end{subequations}
for electrons and holes, respectively. 
In graphene, the electron and hole currents in valley $K$ perpendicular to the interface are \citep{AkhmerovValleyPolarPRL07} 
\begin{subequations}
\begin{align}
j_{K\textrm{gr}}^{e} & =v\left(\psi_{K\textrm{A}}^{e*}\psi_{K\textrm{B}}^{e}+\psi_{K\textrm{B}}^{e*}\psi_{K\textrm{A}}^{e}\right)\\
j_{K\textrm{gr}}^{h} & =-v\left(\psi_{K\textrm{B}}^{h*}\psi_{K\textrm{A}}^{h}+\psi_{K\textrm{A}}^{h*}\psi_{K\textrm{B}}^{h}\right).\label{eq:jhole_Kgr}
\end{align}
\end{subequations} 
The corresponding expressions in the $K'$ valley \citep{AkhmerovValleyPolarPRL07} have a minus sign. 
Note that the current in graphene does not involve the derivative of the wavefunction but rather the amplitudes on the  A and B sites, which are nearest neighbors in the honeycomb lattice. 

There are four cases to consider ($e$ and $h$ in valleys $K$ and
$K'$); as an example, we consider the case of electrons in valley
$K$. Setting the current on the two sides of the interface equal
yields 
\begin{align}
 & 0=j_{S}^{e}-j_{K\textrm{gr}}^{e} \\
 & =\left[\frac{i\hbar}{2m}\psi_{S}^{e}\partial\psi_{S}^{e*}-v\psi_{\textrm{B}}^{e*}\psi_{\textrm{A}}^{e}\right]-\left[\frac{i\hbar}{2m}\psi_{S}^{e*}\partial\psi_{S}^{e}+v\psi_{\textrm{A}}^{e*}\psi_{\textrm{B}}^{e}\right].\nonumber
\end{align}
Writing $\psi_{S}^{e}=|\psi_{S}^{e}|e^{i\theta}$, we divide by $|\psi_{S}^{e}|$
and use the continuity of the wavefunction at the interface to obtain
\begin{equation}
e^{i\theta}\left[\frac{i\hbar}{2m}\partial\psi_{S}^{e*}-v\psi_{\textrm{B}}^{e*}\right]-e^{-i\theta}\left[\frac{i\hbar}{2m}\partial\psi_{S}^{e}+v\psi_{\textrm{B}}^{e}\right]=0.
\end{equation}
Since the overall phase of the wavefunction is arbitrary, this must
hold for any $\theta$, and thus each term must vanish individually, 
yielding the boundary condition Eq.\,(\ref{eq:bndcond2}). 
For holes in the $K$ valley, both $j_{K\textrm{gr}}^{h}$ in (\ref{eq:jhole_Kgr})
and $j_{S}^{h}$ in (\ref{eq:jhole_sup}) acquire a minus sign relative
to the electron case, and so the result is the same. In the $K'$
valley, an additional minus sign is present in the expression for
the current in graphene, yielding a minus sign in the result.
Thus the boundary condition Eq.\,(\ref{eq:bndcond2}) is established.

\section{Continuum Calculation of \\Andreev Reflection at $B=0$ \label{sec:B=0Calc}}

In this appendix we find a scattering wavefunction for an electron
incident from a Dirac material onto a superconductor 
with parabolic dispersion, following the spirit of the classic BTK analysis
\citep{BTK-PRB82}. In the superconductor, the wavefunction is given
by Eq.\,(\ref{eq:psiS}), while in the graphene, a solution in the
$K$ valley has the form (\ref{eq:psiGrK}) where 
$\{\phi_{i}^{\beta}(y)\}$ are incoming and outgoing plane waves. 

At the chemical potential ($E=0$) the graphene wavefunction is ($y<0$) 
\begin{align}
\Psi_{K,k}(y)= & \left[e^{+ik_{y}y}\left(\!\!\begin{array}{c}
1\\
-e^{-i\phi}
\end{array}\!\!\right)+r_{e}e^{-ik_{y}y}\left(\!\begin{array}{c}
1\\
e^{+i\phi}
\end{array}\!\!\right)\right]\!\otimes\!\left(\!\begin{array}{c}
1\\
0
\end{array}\!\right)\nonumber \\
 & +\:r_{h}e^{+ik_{y}y}\left(\!\begin{array}{c}
1\\
-e^{+i\phi}
\end{array}\!\right)\otimes\left(\!\begin{array}{c}
0\\
1
\end{array}\!\right), \label{eq:Psi-Beq0}
\end{align}
where the wavevector perpendicular to the interface is $k_{y}\!\equiv\!\sqrt{k_{F\textrm{gr}}^{2}-k^{2}}\!>\!0$
with $k_{F\textrm{gr}}\!\equiv\!\mu_{\textrm{gr}}/\hbar v$. The 
phase between the A and B sublattices, $\phi\!=\!\arcsin\left(k/k_{F\textrm{gr}}\right)\!\in\![-\pi/2,+\pi/2]$,
depends on the direction of propagation of the plane wave.  From
the boundary conditions Eqs.\,(\ref{eq:k-match})-(\ref{eq:bndcond2}),
the four coefficients $\{r_{e},r_{h},C_{3},C_{4}\}$
are found. The Andreev reflection probability is then given by $P_{A}\equiv|r_{h}|^{2}$.
While we focus on the $E\!=\!0$ case here, note that for $E\neq0$, modifications
are required: $k_{y}$ for the electrons and holes is different, as
is the phase $\phi$
\footnote{In addition, we have suppressed a factor that may be important in
other situations. Strictly speaking, one should consider input and
output beams that all have the same current; thus, the plane waves
in Eq.\,(\ref{eq:Psi-Beq0}) should be divided by $\sqrt{\cos\phi}$. In our situation, different
wavevectors are not mixed, so this is the same for all components
(at $E=0$). If the interface were to cause momentum mixing, then
this effect must be included.}. In the limit $\Delta\ll \mu_{FS}$ and $E<\Delta$
these changes typically introduce only a small change in the result. 

The matching condition for the wavefunction on the A sublattice
yields
\begin{equation}
\left(\!\begin{array}{c}
1+r_e\\
r_h
\end{array}\!\right)=C_{3}\left(\!\begin{array}{c}
1\\
-i
\end{array}\!\right)+C_{4}\left(\!\begin{array}{c}
1\\
+i
\end{array}\!\right),
\end{equation}
while the current continuity condition results in 
\begin{equation}
\left(\!\begin{array}{c}
e^{-i\phi}-r_{e}e^{+i\phi}\\
r_{h}e^{+i\phi}
\end{array}\!\right)=C_{3}\frac{\hbar q}{2mv}\left(\!\begin{array}{c}
1\\
-i
\end{array}\!\right)-C_{4}\frac{\hbar q^{*}}{2mv}\left(\!\begin{array}{c}
1\\
+i
\end{array}\!\right),
\end{equation}
in terms of $q$ defined in Eq.\,(\ref{eq:q_Eeq0}). While the imaginary
part of $q$ is crucial for the sense of the problem (it makes the
wavefunction in the superconductor evanescent), 
it is often not very important in determining the relative amplitude
of the different contributions because $\Delta\lll \mu_{FS}$. So we solve the simplified case
$q=q_{y}$ first. 

\paragraph{Normal Incidence, $q=q_{y}$.}

Initially, we simplify further and consider normal incidence: $k\!=\!0$
and $\phi\!=\!0$. The solution for the Andreev reflection probability 
is Eq.\,(\ref{eq:PA_B0-normal}). 
In view of (\ref{eq:vtildesq1}), the dependence
on the graphene Fermi energy is very weak. This is in striking contrast
to the case of parabolic dispersion: for normal incidence on an interface
between a low-density semiconductor and a superconductor, the result
is Eq.\,(\ref{eq:PA-B0semicond})  
\citep{BTK-PRB82,BlonderTinkhamPRB83,NakanoBTKinterferomPRB93,MortensenAngleAndreevPRB99}. 

\paragraph{Arbitrary Incidence, $q=q_{y}$.}

When not at normal incidence, the solution is
\begin{equation}
P_{A}=\frac{4\widetilde{v}^{2}\cos^{2}\phi}{1+\widetilde{v}^{4}+2\widetilde{v}^{2}\cos(2\phi)}.\label{eq:PA_Beq0}
\end{equation}
Note that for all angles of incidence $\phi$, $P_{A}=1$ when $\widetilde{v}=1$.
The Andreev reflection continues to be smooth, with no strong dependence
on either $q_{y}$ or $\phi$.

\paragraph{Include imaginary part of $q$, $q_{\Delta}^{2}\ll q_{y}^{2}$.}

In most situations, $k_{FS}$ is not close to $K$, and so since $\Delta\lll \mu_{FS}$,
we have $q_{\Delta}^{2}\ll q_{y}^{2}$. Then, using the approximation
Eq.\,(\ref{eq:q_small-Delta-approx}) for $q$, we solve the full
$E=0$ matching equations and expand to leading order in $q_{\Delta}^{2}/q_{y}^{2}$:
\begin{align}
P_{A} & =\frac{4\widetilde{v}^{2}\cos^{2}\phi}{1+\widetilde{v}^{4}+2\widetilde{v}^{2}\cos(2\phi)}
\nonumber \\ & \times
\left[1-\frac{1}{2}\frac{\widetilde{v}^{2}\left(\widetilde{v}^{2}+\cos(2\phi)\right)}{1+\widetilde{v}^{4}+2\widetilde{v}^{2}\cos(2\phi)}\left(\frac{q_{\Delta}^{2}}{q_{y}^{2}}\right)^{2}+\ldots\right].
\end{align}

This correction is very small. For instance, for $\widetilde{v}=1$,
\begin{equation}
P_{A}=\left[1-\frac{1}{4}\left(\frac{q_{\Delta}^{2}}{q_{y}^{2}}\right)^{2}+\ldots\right],
\end{equation}
independent of the angle of incidence, whereas for normal incidence,
\begin{equation}
P_{A}=\frac{4\widetilde{v}^{2}}{\left(1+\widetilde{v}^{2}\right)^{2}}\left[1-\frac{1}{2}\frac{\widetilde{v}^{2}}{1+\widetilde{v}^{2}}\left(\frac{q_{\Delta}^{2}}{q_{y}^{2}}\right)^{2}+\ldots\right].
\end{equation}
These corrections are of order ($\Delta/\mu_{FS})^{2}$ and so can be
neglected.

\paragraph{Include imaginary part of $q$, $q_{\Delta}^{2}$ not small.}

On the other hand, when the Fermi wavevector in the superconductor
is close to matching the valley wavevector in graphene, $k_{FS}\approx K$,
then the transverse wavevector in the superconductor, $q_{y}$, is
small. In that case, the imaginary part of $q$ is not small, and
the approximations in the previous paragraph do not hold. 

We start with the case $q_{y}=0$, implying that $k_{FS}=p=K+k$.
In this case $q^{2}=i\frac{2m}{\hbar^{2}}\Delta$ is pure imaginary,
\begin{equation}
q=\frac{\sqrt{2m\Delta}}{\hbar}e^{i\pi/4}\equiv q_{\Delta}e^{i\pi/4}.
\end{equation}
In terms of the scaled $q_{\Delta}^{2}$ defined by
\begin{align}
\widetilde{q}_{\Delta}^{2}=\left(\frac{\hbar q_{\Delta}}{2mv}\right)^{2} & =\frac{\Delta}{2mv^{2}}\ll1,
\end{align}
the Andreev probability is
\begin{equation}
P_{A}=\frac{2\widetilde{q}_{\Delta}^{2}\cos^{2}\phi}{1+\widetilde{q}_{\Delta}^{4}+2\widetilde{q}_{\Delta}^{2}\cos(2\phi)}\approx2\widetilde{q}_{\Delta}^{2}\cos^{2}\phi\ll1.
\end{equation}
Thus the Andreev reflection is very small in this limit, of order
$\Delta/\mu_{FS}$.

Now suppose that $q_{y}$ is of order $q_{\Delta}$. Then \emph{both}
$\widetilde{v}^{2}$ and $\widetilde{q}_{\Delta}^{2}$ are small.
The result will be proportional to either $\widetilde{v}^{2}$ or
$\widetilde{q}_{\Delta}^{2}$, implying that Andreev reflection is
weak whenever the imaginary part of $q$ is significant.

\paragraph{Summary.}

Thus we see that the simple results Eqs.\,(\ref{eq:PA_B0-normal}) and
(\ref{eq:PA_Beq0}) provide a good guide to the strength of Andreev
reflection, unless $q_{y}\lesssim q_{\Delta}$. The latter condition
in terms of energy is $\mu_{S}-\hbar^{2}(K+k)^{2}/2m\lesssim\Delta$:
we see that for special fine-tuned values of $\mu_{S}$ the Andreev
reflection is small. However, in typical situations, we can expect
good Andreev reflection at clean graphene-superconductor interfaces.



\section{Chiral Andreev Edge States in the Continuum Model \label{sec:CAES-Calc}}

In this appendix we show how we obtain 
the interface modes in the continuum model. First, we fix the form of the transverse wavefunction. It is convenient to introduce the following shifted coordinates and scaled energies: 
\begin{align}
y_{\pm}\equiv\sqrt{2}\left(y\pm kl_{B}^{2}\right),\quad & \alpha_{\pm}=\frac{\mu_{\textrm{gr}}\pm E}{E_{L}},
\end{align}
where the plus sign pertains to electrons and minus 
to holes. 
The functions in Eq.\,(\ref{eq:psiGrK})
are \citep{HoppeZulickePRL00,AkhmerovValleyPolarPRL07,Romanovsky_DiracQH_PRB11}
\begin{align}
\left(\!\begin{array}{c}
\phi_{1k}^{e}(y)\\
\phi_{2k}^{e}(y)
\end{array}\!\right) & =\frac{N_{e}}{\sqrt{l_{B}}}\left(\!\begin{array}{c}
D_{\alpha_{+}^{2}}(-y_{+}/l_{B})\\
-i\alpha_{+}D_{\alpha_{+}^{2}-1}(-y_{+}/l_{B})
\end{array}\!\right),\\[0.1cm]
\left(\!\begin{array}{c}
\phi_{1k}^{h}(y)\\
\phi_{2k}^{h}(y)
\end{array}\!\right) & =\frac{N_{h}}{\sqrt{l_{B}}}\left(\!\begin{array}{c}
-i\alpha_{-}D_{\alpha_{-}^{2}-1}(-y_{-}/l_{B})\\
D_{\alpha_{-}^{2}}(-y_{-}/l_{B})
\end{array}\!\right),
\end{align}
where the $D_{n}(x)$ are parabolic cylinder functions \citep{ParCylFun_NIST10}
 and $N_{e,h}$ are normalization coefficients determined by Eq.\,(\ref{eq:Norm-phis}).

The boundary conditions Eqs.\,(\ref{eq:bndcond1})-(\ref{eq:bndcond2})
fix the ratio of the coefficients in the transverse wavefunction.
It is convenient to use a scaled wavevector and its real and imaginary
parts,
\begin{align}
\widetilde{q} & \equiv\frac{\hbar q}{2mv}=\widetilde{q}'+i\widetilde{q}'',
\end{align}
where $q$ is defined in Eq.\,(\ref{eq:q_Eeq0}). Note that $\widetilde{q}^{2}=\widetilde{v}^{2}+i\Delta/2mv^{2}$,
but that $\widetilde{q}'\approx\widetilde{v}$ only in the approximation
$\Delta\ll\frac{1}{2}mv_{y}^{2}$. The boundary conditions then lead to 
\begin{align}
\frac{C_{1}}{C_{4}} & =\sqrt{\frac{l_{B}}{\xi_{y}}}\frac{2\widetilde{q}'}{\widetilde{q}\phi_{1k}^{e}(0)+\phi_{2k}^{e}(0)},\label{eq:C1dC4}\\[0.2cm]
\frac{C_{2}}{C_{4}} & =\gamma^{*}\sqrt{\frac{l_{B}}{\xi_{y}}}\frac{2\widetilde{q}'}{\phi_{1k}^{h}(0)\widetilde{q}+\phi_{2k}^{h}(0)},\label{eq:C2dC4}\\[0.2cm]
\frac{C_{3}}{C_{4}} & =\exp\left[-2i\arg\left\{ \frac{\widetilde{q}}{\phi_{2k}^{e}(0)}+\frac{1}{\phi_{1k}^{e}(0)}\right\} \right]\label{eq:C3dC4a}\\
 & =-\exp\left[-2i\arg\left\{ \frac{\widetilde{q}\gamma}{\phi_{2k}^{h}(0)}+\frac{\gamma}{\phi_{1k}^{h}(0)}\right\} \right].\label{eq:C3dC4b}
\end{align}
The first two equations imply that the ratio of electron to hole content
in graphene is
\begin{equation}
\frac{C_{1}}{C_{2}}=\gamma\frac{\widetilde{q}\phi_{1k}^{h}(0)+\phi_{2k}^{h}(0)}{\widetilde{q}\phi_{1k}^{e}(0)+\phi_{2k}^{e}(0)}\label{eq:AppA_C1dC2}
\end{equation}
for the $K$-valley mode. 

\subsection{Secular equation}
\label{subsec:SecularEq}

The secular equation for the spectrum follows from the equality of
the electron and hole expressions for $C_{3}/C_{4}$ [Eqs.\,(\ref{eq:C3dC4a})-(\ref{eq:C3dC4b})]. It is convenient
to define the ratios 
\begin{align}
f_{\pm}(k) & \equiv\frac{D_{\alpha_{\pm}^{2}}\left(\mp\sqrt{2}kl_{B}\right)}{\alpha_{\pm}D_{\alpha_{\pm}^{2}-1}\left(\mp\sqrt{2}kl_{B}\right)}.
\end{align}
The expressions above for $C_{3}/C_{4}$ in the $K$ valley lead to
\begin{align}
|\widetilde{q}|^{2}f_{+}-f_{-} & =\frac{E}{\sqrt{\Delta^{2}-E^{2}}}\left(f_{+}f_{-}+1\right)\widetilde{q}'-\left(f_{+}f_{-}-1\right)\widetilde{q}''.\label{eq:f+f-_forK}
\end{align}

\begin{figure}
\includegraphics[width=2.3in]{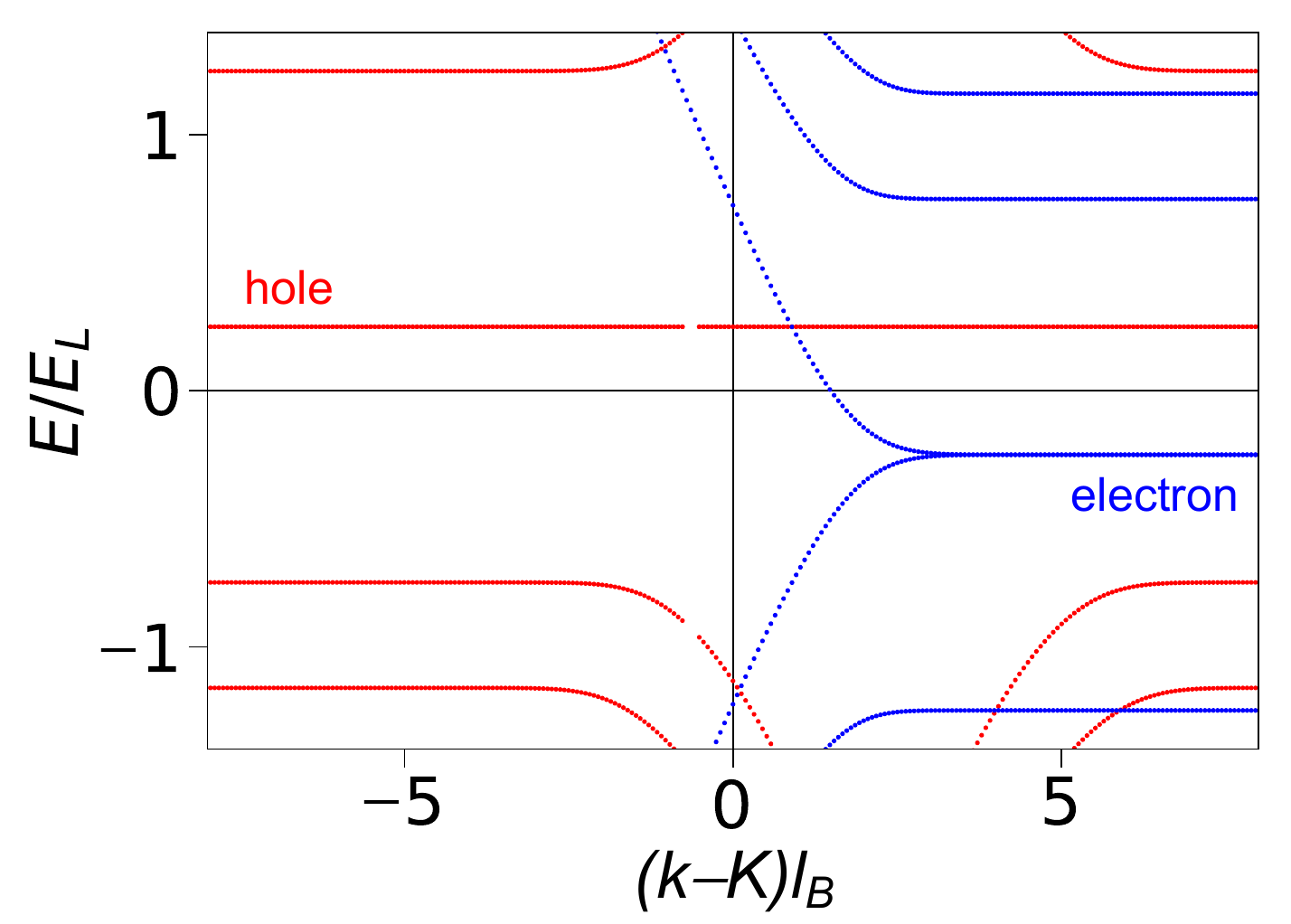}
\caption{Spectrum of electron (blue) and hole
(red) states for a zigzag nanoribbon in the QH regime. Zoom on the $K$-valley of the tight-binding results in Fig.\,\ref{fig:spectrum_term-latt}(c): $k$ is scaled
by $l_{B}$ for direct comparison to the continuum results in Fig.\,\ref{fig:spectrum_term-latt}(a). 
(Standard parameters, Sec.\,\ref{subsec:Parameters}.)}
\label{fig:spec_term-latt_zoom}
\end{figure}

By comparing the form of the wavefunction in the $K$ and $K'$ valleys,
one finds that the corresponding equation for the $K'$ valley is
obtained by the substitution $f_{\pm}\longrightarrow-1/f_{\pm}$.
The same equations are obtained by the substitution $k\longrightarrow-k$
and $E\longrightarrow-E$ in (\ref{eq:f+f-_forK}), thus verifying
the inherent BdG particle-hole symmetry. 

The normal QH edge states (Figs.\,\ref{fig:termin-latt-psi} and \ref{fig:spectrum_term-latt}) can be readily obtained by considering a
large barrier at the interface, $V_{0}\to\infty$ in Eqs.\,(\ref{eq:def_V0})
and (\ref{eq:V0-bndcond}). In that case, (\ref{eq:f+f-_forK}) yields
$f_{+}=0$, which indeed reduces to the usual condition for a QH edge
state. As an example of the good quantitative agreement between the continuum and tight-binding models for the QH edge states, in Fig.\,\ref{fig:spec_term-latt_zoom} we show a zoom of the $K$ valley tight-binding spectrum with $k$ scaled by $l_B$ to facilitate comparison with Fig.\,\ref{fig:spectrum_term-latt}(a).

\subsection{Momentum at $E=0$ }

\begin{figure}[t]
\includegraphics[width=2.5in]{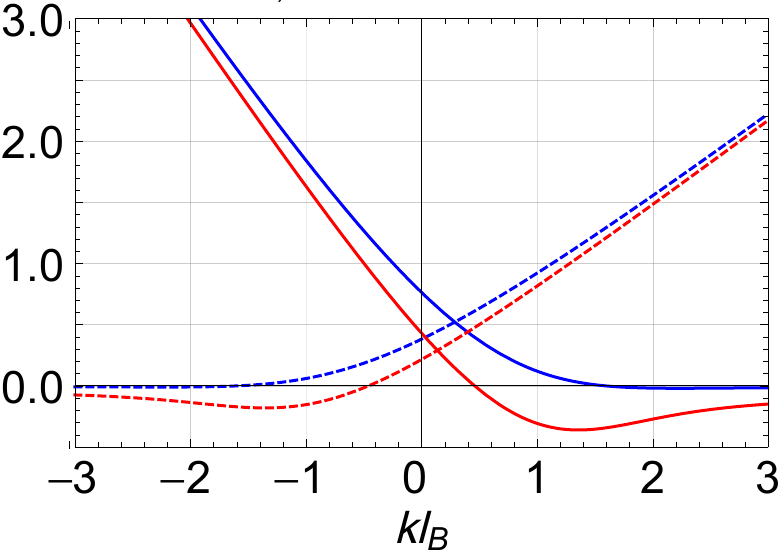}
\caption{Graphical solution of the transcendental
equation $\alpha f_{-}=\alpha\widetilde{v}^{2}f_{+}$ for the momentum
of the $E=0$ CAEM: dashed (solid) line for left-hand (right-hand)
side. Results for two chemical potentials are shown, $\alpha\!=\!\mu_{\textrm{gr}}/E_{L}\!=\!0.25$
(blue) and $0.75$ (red). Here, $\widetilde{v}^{2}\!=\!2$ and the imaginary
part of $\widetilde{q}$ is neglected.}
\label{fig:SolvingContinuum}
\end{figure}

Taking $E=0$, we have $\alpha_{+}=\alpha_{-}\equiv\alpha$
and $f_{+}(k_{0})=f_{-}(-k_{0})$. 
Consider the graphical solution of the secular equation in a very simple limit:
ignore the imaginary part of $\widetilde{q}$ completely, $\widetilde{q}''=0$.
In this case, the secular equation reduces to 
\begin{equation}
f_{-}(k_{0})=\widetilde{v}^{2}f_{+}(k_{0}).\label{eq:Simple_SecEq}
\end{equation}
The two sides of this equation (multiplied by the graphene doping) are plotted in Fig.\,\ref{fig:SolvingContinuum}
 for $\widetilde{v}^{2}=2$ and two values of $\alpha$: the right-hand
side is simply the mirror image of the left-hand side rescaled by
$\widetilde{v}^{2}$.

The factor $\widetilde{v}^{2}$ provides the asymmetry between the
electron and hole branches and generally shifts the momentum away
from the valley. When $\widetilde{v}^{2}=1$, the curves are mirror
images and so intersect at $k_{0}=0$ for all $\alpha$. This is the
valley-degenerate case. On the other hand, when $\alpha=1$ (the top
of the LLL), the curves go through the origin, $f_{\pm}(0)=0$, and
so necessarily intersect there: $k_{0}=0$ for all $\widetilde{v}^{2}$
at the top of the LLL. Whether $\widetilde{v}^{2}\lessgtr1$ determines
the direction of the momentum shift, $k_{0}\lessgtr0$. Finally, the
largest possible momentum with $\alpha$ fixed 
is given by the uncoupled, terminated-lattice case $\widetilde{v}^{2}\to\infty$. 

\begin{figure}
\includegraphics[width=2.7in]{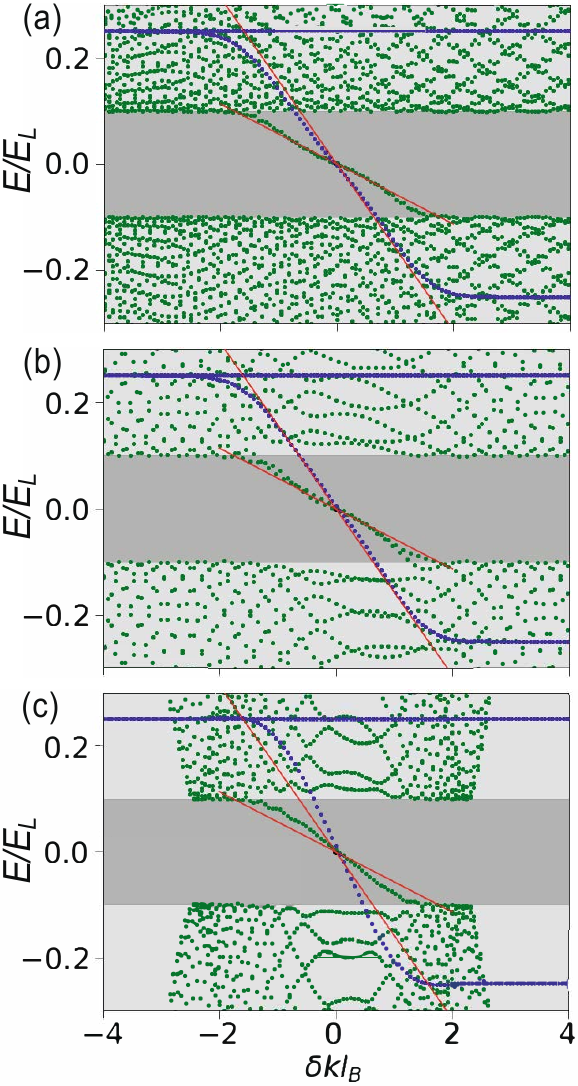}
\caption{Spectra near the $K$ point for three
transparent tight-binding interfaces---(a) dense, (b) sparse, and
(c) honeycomb---compared to the group velocity of the continuum model
(red straight lines). Results for two values of the superconducting
gap are superposed, $\Delta/E_{L}=0.1$ (green) and $0.5$ (blue).
The continuum model is an excellent description of the tight-binding results. 
(Standard parameters, Sec.\,\ref{subsec:Parameters}.)}
\label{fig:modevelocities}
\end{figure}

\subsection{Group velocity \label{subsec:Vg_appendix}}

The group velocity of the spectrum at $E=0$ has a simple explicit
form if one continues to neglect the imaginary part of $\widetilde{q}$.
Implicit differentiation of the secular equation, Eq.\,(\ref{eq:Simple_SecEq}),
yields 
\begin{align}
\hbar v_{g}= & \frac{\partial E}{\partial k}=\sqrt{2}\mu_{\textrm{gr}}l_{B}\frac{\frac{1}{\widetilde{v}}+\widetilde{v}}{\lambda\left(\alpha,k_{0},\widetilde{v}\right)+E_{L}/\Delta},
\end{align}
where the auxiliary function $\lambda$ is defined by 
\begin{align}
\lambda(\alpha,k_{0},\widetilde{v})\equiv & -\frac{w(\alpha,z_{0})/\widetilde{v}+w(\alpha,-z_{0})\widetilde{v}}{1+f_{+}(k_{0})f_{-}(k_{0})}
\end{align}
for $z\equiv\sqrt{2}kl_{B}$ and 
\begin{align}
w(\alpha,z)\equiv & -\frac{f_{+}}{\alpha}+2\frac{1}{D_{\alpha^{2}-1}\left(-z\right)}\left.\partial_{\nu}D_{\nu}\left(-z\right)\right|_{\nu=\alpha^{2}}\nonumber \\
 & -2\frac{D_{\alpha^{2}}\left(-z\right)}{D_{\alpha^{2}-1}^{2}\left(-z\right)}\left.\partial_{\nu}D_{\nu}\left(-z\right)\right|_{\nu=\alpha^{2}-1}\;.
\end{align}

We compare these continuum model results to tight-binding calculations
for a transparent, well-matched interface in Fig.\,\ref{fig:modevelocities}. Despite
taking $\widetilde{q}''=0$ here, the agreement 
for dense and sparse stitching is excellent. 

We can estimate the change in $v_{g}$ between the QH edge states
and the fully hybridized CAEM by comparing two limits. $\lambda(\alpha,k_{0},\widetilde{v})$
needs to be evaluated at the edge state momentum. However, at good
coupling this function is basically flat: $\lambda(\alpha,0,1)\in[2.5,5]$
from $\alpha=0$ to $1$. In the disconnected case, $\widetilde{v}\rightarrow\infty$ 
implies $\lambda(\alpha,k_0,\widetilde{v})/\widetilde{v}\rightarrow-w(\alpha,-z_{c})\sim2\alpha+1/2$.
Taking the ratio yields 
\begin{align}
\frac{v_{g,\textrm{CAEM}}}{v_{g,\textrm{QH}}} & 
=\frac{-2w(\alpha,-z_{0})}{\lambda(\alpha,0,1)+E_{L}/\Delta}
\sim\frac{1+4\mu_{\textrm{gr}}/E_{L}}{\lambda(\alpha,0,1)+E_{L}/\Delta}.
\end{align}
Thus we see that the Andreev edge modes in the typical case $E_L \gg \Delta$
are substantially slower. For strong magnetic quantization ($E_{L}/\Delta\gg\lambda$),
a good estimate of the CAEM group velocity is 
\begin{equation}
\frac{\hbar v_{g,\textrm{CAEM}}}{l_B}\sim\sqrt{2}\Delta\left(\frac{1}{|\widetilde{q}|}+|\widetilde{q}|\right)\frac{\mu_{\textrm{gr}}}{E_{L}},
\end{equation}
 proportional to the pairing gap and a factor characterizing interface
quality. 

\bibliography{../../QTransport_2025-06,BibFootnotes} 

\begin{thebibliography}{103}
\bibitem [{Dis()}]{Discont-Fig1c}%
  \BibitemOpen
  \href@noop {} {}\bibinfo {note} {The solutions in the lower left and upper
  right of Fig.\,\ref{fig:MainResults}(c) are at the energy of the
  superconducting gap. They are artifacts of the sharp edge of the
  superconductor's spectrum and so not relevant for our discussion of the
  CAEM.}\BibitemShut {Stop}%
\bibitem [{v-p()}]{v-parabolic}%
  \BibitemOpen
  \href@noop {} {}\bibinfo {note} {It is possible for two materials with
  parabolic dispersion to have the same Fermi velocity, $v_{F1}=v_{F2}$,
  despite having very different densities, $k_{F1} \ll k_{F2}$, so that
  material 1 is a tunable semiconductor and material 2 is a superconductor.
  However, this requires the effective masses to be very different, $m_2/m_1
  \sim 100$. In this case, the probability of Andreev reflection at $B=0$ could
  be of order $1$ [Eq.\,(\ref{eq:PA-B0semicond})], and we believe that Andreev
  reflection in the QH regime would also be strong (but we are not aware of a
  calculation). The key point is that the velocities in the two materials be
  similar so that current continuity is easily satisfied. Because we don't know
  of two such materials, we think this scenario unlikely and do not consider it
  further.}\BibitemShut {Stop}%
\bibitem [{Wid()}]{WidthFootnote}%
  \BibitemOpen
  \href@noop {} {}\bibinfo {note} {The width of the zigzag graphene nanoribbon
  is $y_{QH}=\frac{300}{\sqrt{3}}a$. The width of the superconductor is,
  respectively, $y_S= 150a$, $400a$, and $\frac{300}{\sqrt{3}}a$ for our dense,
  sparse, and honeycomb results.}\BibitemShut {Stop}%
\bibitem [{BdG()}]{BdG-redundancy}%
  \BibitemOpen
  \href@noop {} {}\bibinfo {note} {In looking at such BdG spectra one must keep
  in mind the inherent redundancy due to the doubling of the degrees of
  freedom.}\BibitemShut {Stop}%
\bibitem [{fn:()}]{fn:SpinConvention}%
  \BibitemOpen
  \href@noop {} {}\bibinfo {note} {Our convention for labeling the spin of the
  Andreev reflected hole follows that for the linear momentum: an incoming spin
  up particle forms a Cooper pair with a spin down particle. The missing spin
  down angular momentum implies that the Andreev reflected hole is spin up.
  Thus an incoming electron Andreev reflects into a hole in the same valley
  with the same spin \cite{ColemanBook}.}\BibitemShut {Stop}%
\end{thebibliography}

\end{document}